\documentclass[a4paper, 10pt, twocolumn,nofootinbib,superscriptaddress]{revtex4-2}
\usepackage{graphicx} 
\usepackage{xcolor} 

\usepackage{amsmath, amssymb, amsfonts}
\usepackage{hyperref}
\hypersetup{colorlinks = true}
\usepackage{physics}
\usepackage{blindtext}
\usepackage{tikz}
\usetikzlibrary{decorations.markings, arrows.meta, patterns}
\usepackage{caption, subcaption}
\usepackage{cancel}
\usepackage[normalem]{ulem}

\definecolor{Acol}{RGB}{244,199,199}
\definecolor{Bcol}{RGB}{200,221,242}
\definecolor{Ccol}{RGB}{205,232,205}
\definecolor{Dcol}{RGB}{250,238,200}

\newcommand{\NTHU}{Department of Physics, National Tsing Hua University, Hsinchu 300044, Taiwan}
\newcommand{\YITP}{Yukawa Institute for Theoretical Physics, Kyoto University, Kyoto 606-8502, Japan}
\newcommand{\maizuru}{National Institute of Technology, Maizuru College, Kyoto 625-8511, Japan}

\newcommand\GM[1][]{\if\relax\detokenize{#1}\relax{\rm GM}\else{\rm GM}^{(#1)}\fi}
\renewcommand\S[1][]{\if\relax\detokenize{#1}\relax S\else S^{(#1)}\fi}
\newcommand\Z[1][]{\if\relax\detokenize{#1}\relax Z\else Z^{(#1)}\fi}

\begin{document}

\title{Genuine Multi-Entropy in the Toric Code}
\author{Sriram Akella}
\email{sriram.akella@tifr.res.in}
\affiliation{Department of Theoretical Physics, Tata Institute of Fundamental Research, Homi Bhabha Road, Mumbai, 400005, India.}

\author{Norihiro Iizuka}
\email{iizuka@phys.nthu.edu.tw}
\affiliation{\NTHU}
\affiliation{\YITP}

\author{Akihiro Miyata}
\email{akihiro.miyata@yukawa.kyoto-u.ac.jp}
\affiliation{\YITP}
\affiliation{\maizuru}

\date{\today}

\begin{abstract} 
We study genuine multi-entropy as a diagnostic of multipartite entanglement in the toric code, which provides a controlled setting for probing multipartite structures in topologically ordered states. Our main question is whether genuine multi-entropy captures information that is not reducible to conventional lower-party entropic data, such as topological entanglement entropy. We first analyze toric-code ground states that admit a stabilizer-state description, where the relevant quantities can be evaluated exactly. We conjecture that, for stabilizer states and $\mathtt{q}\ge4$, the $\mathtt{q}$-partite genuine multi-entropy at replica index $n<\mathtt{q}$ collapses to a linear combination of multi-entropies involving at most $\mathtt{q}-2$ parties. We establish this pattern explicitly for $\mathtt{q}=4$ in the toric code stabilizer sector: for $n=2,3$, the genuine multi-entropy is proportional to the tripartite information $I_3$ and, for the Kitaev--Preskill partition, contains no independent genuine four-partite information beyond that captured by the topological entanglement entropy. At $n=4$, however, this reduction breaks down: the genuine multi-entropy is no longer proportional to $I_3$, but remains a topological invariant of the toric-code stabilizer ground states under Kitaev--Preskill-preserving deformations that do not change whether the numbers of vertex- and plaquette-type junctions coincide. For generic non-stabilizer superpositions within the ground-state manifold and for coherent superpositions of local excitations, the low-$n$ reduction also fails. These results show that genuine multi-entropy probes multipartite entanglement structure beyond the tripartite information, and hence beyond the topological entanglement entropy in the Kitaev--Preskill partition, whereas for stabilizer states at low replica index it reduces to lower-partite entropic data.
\end{abstract}

\maketitle

\section{Introduction and summary}
\label{Intro}

Topologically ordered phases of matter \cite{Wen:2012hm, Wen:2016ddy} are characterized by long-range entanglement in their ground-state wavefunctions and cannot be described by conventional local order parameters. Entanglement therefore provides a natural language for probing their topological structure. A central example is the topological entanglement entropy (TEE) \cite{Kitaev:2005dm, Levin:2006zz}, which is constructed from a suitable linear combination of \emph{bipartite} entanglement entropies so as to cancel nonuniversal boundary contributions and isolate universal topological information. The role of genuinely \emph{multipartite} entanglement in concrete lattice realizations of topological order, however, remains much less explored.

In this work we study genuine multipartite entanglement in Kitaev’s toric code \cite{Kitaev:1997wr}, using the genuine multi-entropy (GM) \cite{Iizuka:2025ioc,Iizuka:2025caq} as a diagnostic. The genuine multi-entropy, GM, is built from the multi-entropy \cite{Gadde:2022cqi} by a genuineness prescription that subtracts lower-partite contributions, so that GM isolates entanglement that is intrinsically shared among all parties. More recently, a general mathematical framework for genuine multipartite
entanglement signals has been formulated in \cite{Gadde:2026msg}. While these quantities are well defined mathematically, their physical interpretation is far from being fully understood. In particular, it remains unclear what information GM captures in topologically ordered states and how it responds to stabilizer structure and local excitations.

To address these questions, we analyze four-partite partitions of the toric-code lattice on the torus and compute GM for (i) the stabilizer ground states, (ii) general ground-state superpositions within the topological ground-state manifold, which need not themselves be stabilizer states, and (iii) excited states. The toric code is particularly well suited for this purpose: it is among the simplest exactly solvable lattice realizations of a topologically ordered phase, its finite-size ground-state wavefunctions and multipartite partitions can be treated explicitly, and on a torus its ground-state manifold contains a distinguished basis of stabilizer states while generic superpositions within the same manifold do not. It therefore allows us to distinguish features of GM that arise from stabilizer structure from those that persist throughout the topological ground-state sector.

Before turning to the detailed analysis, we first explain the broader
motivation from the relation between multipartite entanglement and topological quantum field theory (TQFT),
then describe the role of stabilizer structure in the toric code, and
finally summarize our main results.

\paragraph*{Broader motivation.}
A broader motivation for this work comes from the relation between entanglement and TQFT. The low-energy effective  description of Levin and Wen’s string-net models \cite{Levin:2004mi} is a TQFT, and a natural question is how to extract TQFT data from the ground-state wavefunction. A celebrated example is the TEE \cite{Levin:2006zz,Kitaev:2005dm}, which extracts the total quantum dimension
$\mathcal{D}$
of the underlying TQFT.  Ground-state entanglement also encodes the modular $S$ and $T$ matrices \cite{Zhang:2011jd}, and more recently multipartite entanglement of the ground state has been used to extract finer TQFT data \cite{Sheffer:2025jtc,DelZotto:2026fpw}. In particular, \cite{DelZotto:2026fpw} relate genuine multipartite entanglement signals to TQFT partition functions on {graph-encoded manifolds} (gems) \cite{ferri1986graph}, up to some corrections that vanish in the continuum limit. 

\paragraph*{Stabilizer structure.}
The toric code \cite{Kitaev:1997wr} is simultaneously a topological phase and a stabilizer quantum error-correcting code \cite{Gottesman:1997zz}: its energy eigenspaces admit bases of stabilizer states. We use the framework of \cite{Akella:2026xza} for evaluating multi-invariants \cite{Gadde:2024taa}, and the associated measures of multipartite entanglement \cite{Ma:2023ecg,Guhne:2008qic}, on stabilizer states. This lets us compute GM explicitly for stabilizer eigenstates of the toric code and ask in what sense a genuinely multipartite signal captures underlying TQFT data.

\paragraph*{Main results.}
One of our main conclusions is the existence of a threshold in the R\'enyi index $n$ for the GM of the stabilizer states.
As we will review more in the next section, GM depends on a R\'enyi index $n$, which is associated with the replica number. 
We conjecture that, for generic $\mathtt{q} \ge 4$, the $\mathtt{q}$-partite GM of a stabilizer state at replica index $n<\mathtt{q}$ collapses to a linear combination of multi-entropies involving at most $\mathtt{q}-2$ parties, with $n=\mathtt{q}$ the threshold at which this reduction no longer holds. We establish this pattern explicitly for the $\mathtt{q}=4$ case in the toric-code examples studied here. For $n=2, 3$, the four-partite GM of a stabilizer state collapses to the R\'enyi tripartite information $I_{3,n}$, 
\begin{align}
&I_{3,n}
=
S^{(2)}_n(BCD:A)
+S^{(2)}_n(CDA:B)
\nonumber\\
&\qquad \qquad
+S^{(2)}_n(DAB:C)
+S^{(2)}_n(ABC:D)
\label{eq:I3n_definition}
\\
&
-S^{(2)}_n(AB:CD)
-S^{(2)}_n(AC:BD)
-S^{(2)}_n(AD:BC). \nonumber
\end{align}
Here the superscript $(2)$ indicates the bipartite $\mathtt{q}=2$
R\'enyi entropy, while the subscript $n$ denotes the
R\'enyi index. Note that every term in $I_{3,n}$ is itself a \emph{bipartite} entanglement entropy; in the multi-entropy counting of Sec.~\ref{review}, such quantities are classified as $\mathtt{q}=2$.

At $n=2$ one has
\begin{align}
\label{eq:GMn2}
 \GM[\mathtt{q}=4]_{n=2} (A:B:C:D) = - \left(a - \tfrac{1}{12}\right) I_{3} ,
 \end{align}
where $I_3$ denotes the common value of the R\'enyi tripartite information $I_{3,n}$ for stabilizer states, which is independent of $n$.
$a$ is the convention parameter of $\GM[\mathtt{q}=4]_n$ \cite{Iizuka:2025ioc}. In particular, with the standard choice $a=\tfrac{1}{3}$ this gives $\GM[\mathtt{q}=4]_{n=2}=-\frac{1}{4} I_3$, which was first conjectured in \cite{Iizuka:2025pqq}. 
The reduction at $n=2$ is special: the corresponding multi-entropy is a Coxeter multi-invariant and reduces, for stabilizer states, to a combination of bipartite entanglement entropies \cite{Akella:2026xza}.

Remarkably, an analogous reduction persists at $n=3$:
\begin{equation}
\label{eq:GMn3}
\GM[\mathtt{q}=4]_{n=3}(A:B:C:D)
=
-\left(a-\tfrac{1}{9}\right) I_3.
\end{equation}
Unlike the $n=2$ relation, this reduction is not a consequence
of the Coxeter structure of the corresponding multi-entropy.
It is therefore a nontrivial property of the stabilizer sector.

Note that for a Kitaev--Preskill disk partition, where $A$, $B$, and $C$
form a disk and $D$ is the exterior region \cite{Kitaev:2005dm}, the tripartite
information $I_{3}$ is the negative of the topological entanglement entropy,
\begin{equation}
I_{3}=-\gamma=-\log\mathcal{D} = - \log  \sqrt{\textstyle\sum_i d_i^2} \,.
\end{equation}
Thus, the collapse of GM to $I_3$ for stabilizer states in $n=2$ and $n=3$ means that, for $n < 4$,
GM is completely determined by the same total-quantum-dimension
data as the TEE, {\it i.e.}, GM contains no independent genuinely four-partite information beyond the TEE for the Kitaev--Preskill disk partition.

At $n=4$, by contrast, this reduction to $I_3$ no longer
holds. The GM then contains information beyond the tripartite
information and hence beyond the total quantum dimension
$\mathcal{D}$ probed by $I_3$. For the toric code stabilizer states, we argue that the $n = 4$ GM is a topological invariant in the sense that it remains unchanged under a precisely defined class of deformations of the partitions (Sec.~\ref{KitaevPreskillpartitions}), and is independent of the lattice size $L$. Moreover, we explicitly show that the $n = 4$ GM does not collapse to $I_3$ by constructing two partitions of the $3 \times 3$ lattice with different ratios of $\text{GM}/I_3$. As a consistency check, we also evaluate the $n = 4$ GM for the 2D color code \cite{Bombin:2006sj, Bombin:2006rc}, whose underlying TQFT is equivalent to two copies of the toric code TQFT \cite{Bombin:2011qp}. Both $I_3$ and the $n = 4$ GM double accordingly, providing further evidence that the $n = 4$ GM is indeed a topological invariant.

Finally, we show that the reductions \eqref{eq:GMn2} and
\eqref{eq:GMn3} are special properties of the stabilizer sector. We show explicitly that, for
generic non-stabilizer states within the same topological ground state manifold, and for excited states obtained by coherent superpositions of local excitations, 
these relations no longer hold, so that
GM at generic R\'enyi index $n$, including $n=2$ and $n=3$, is no longer
captured by $I_{3,n}$ alone. In this sense GM detects more than the total quantum
dimension $\mathcal{D}$ throughout the ground-state manifold; its reduction
to lower-partite TEE data is special to the stabilizer states. 

This paper is organized as follows. In Sec.~\ref{review}, we review GM, the
toric code, and other relevant background material. Section~\ref{StabilizerGM}
presents one of our main results: a study of GM for stabilizer states of the
toric code. Section~\ref{sec:beyond-stabilizer} studies GM beyond the
stabilizer ground-state sector, and Sec.~\ref{sec:local-excitations} studies GM for excited states. We comment on future directions in
Sec.~\ref{discussions}, and collect several technical details in the
appendices.

\section{Genuine multi-entropy (GM) and Toric codes}
\label{review}

\subsection{Genuine multi-entropy (GM)}
\label{sec:GM_review}

We briefly review the R\'enyi genuine multi-entropy introduced in
Refs.~\cite{Iizuka:2025ioc,Iizuka:2025caq}.  The R\'enyi-$n$
$\mathtt{q}$-partite multi-entropy $S_n^{(\mathtt{q})}(A_1:\cdots:A_\mathtt{q})$ is defined by a
symmetric replica contraction of $n^{\mathtt{q}-1}$ copies of the state
\cite{Gadde:2022cqi,Penington:2022dhr,Gadde:2023zzj}.
For $\mathtt{q}=2$, it reduces to the ordinary R\'enyi entropy.  For $\mathtt{q}>2$,
however, $S_n^{(\mathtt{q})}$ receives contributions not only from irreducible
$\mathtt{q}$-partite correlations but also from lower-partite correlations.

The genuine multi-entropy is constructed by taking suitable linear
combinations of multi-entropies so as to subtract these lower-partite
contributions.  It therefore provides a diagnostic for irreducible
multipartite entanglement.  In this work, we focus on the 
four-partite quantity $\GM[\mathtt{q}=4]$,
\begin{align}
&\GM[4]_n(A:B:C:D)
=
S^{(4)}_n(A:B:C:D)
\nonumber\\
&
\hspace{-3mm}-\frac{1}{3}\Big[
S^{(3)}_n(AB:C:D)
+S^{(3)}_n(AC:B:D)
+S^{(3)}_n(AD:B:C)
\nonumber\\
&
+S^{(3)}_n(BC:A:D)
+S^{(3)}_n(BD:A:C)
+S^{(3)}_n(CD:A:B)
\Big]
\nonumber\\
&
+\frac{1}{3}\Big[
S^{(2)}_n(ABC:D)
+S^{(2)}_n(ABD:C)
+S^{(2)}_n(ACD:B) \nonumber\\
& \hspace{10mm}
+S^{(2)}_n(BCD:A)
\Big]
-a\, I_{3,n}.
\label{eq:GM4_definition}
\end{align}
The last term is proportional to the R\'enyi tripartite information \eqref{eq:I3n_definition} and $a$ is a real parameter. 

Thus, different choices of $a$ differ only by an $I_{3,n}$ contribution.
Since we will investigate the dependence of the result on this
ambiguity, we keep $a$ arbitrary throughout this work.

For completeness, the replica definition of $S_n^{(\mathtt{q})}$ is given in
Appendix~\ref{app:Mdefinition}.

\subsection{Toric codes and Kitaev--Preskill partitions}
\label{KitaevPreskillpartitions}

\begin{figure}
    \centering 
    \includegraphics[width=\linewidth]{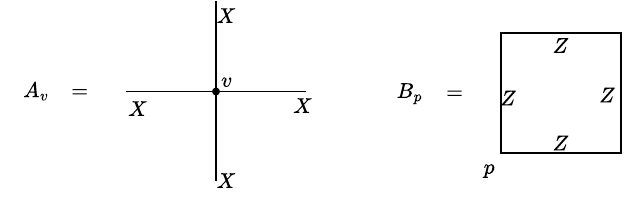}
    \caption{Star and plaquette operators in the toric code.}
    \label{fig:star-plaq}
\end{figure}

The toric code is defined on a lattice embedded in a closed manifold, with spin-$1/2$ degrees of freedom living on the edges. 
The Hamiltonian of the system is
\begin{equation}
 H = - \sum_{\text{vertices}} A_v - \sum_{\text{plaquettes}} B_p,
\end{equation}
where $A_v$ and $B_p$ are the star and plaquette operators. $A_v$ acts with the Pauli $X$ on the incident edges of a vertex $v$, denoted as $+_v$, and $B_p$ acts with a Pauli $Z$ on the plaquettes $p$ of the lattice, denoted as $\square_p$, thus 
\begin{equation}
A_v = \prod_{i \in +_v} X_i \,, \qquad 
B_p = \prod_{i \in \square_p} Z_i .
\end{equation}
The star and plaquette operators are shown in Fig. \ref{fig:star-plaq}. 
The star and plaquette operators commute with each other, and the ground states of the above Hamiltonian are simultaneous eigenstates of all the $A_v$ and $B_p$ operators, 
\begin{equation}
A_v \ket{\psi} = \ket{\psi} \,, \quad
B_p \ket{\psi} = \ket{\psi}
\end{equation}

Since the spatial manifold is closed, the star and plaquette operators satisfy
\begin{equation}
 \prod_{v} A_v = \prod_{p} B_p = 1,
\end{equation}
where the product is over all vertices and all plaquettes respectively. Thus, among the $N_v+N_p$ stabilizer constraints, only $N_v+N_p-2$ are independent. Since the number of physical qubits is $N_e$, the number of logical qubits in the ground-state subspace is
\begin{equation}
N_e-(N_v+N_p-2)
= 2-\chi,
\end{equation}
where
\begin{equation}
\chi=N_v-N_e+N_p
\end{equation}
is the Euler characteristic of the spatial manifold. Therefore, the ground-state degeneracy $N_{gsd}$ is
\begin{equation}
N_{gsd}=2^{2-\chi} = 2^{2g}=4^g .
\end{equation}
where $g$ is the genus of the manifold.

\begin{figure}
    \centering 
\includegraphics[width=0.9\linewidth]{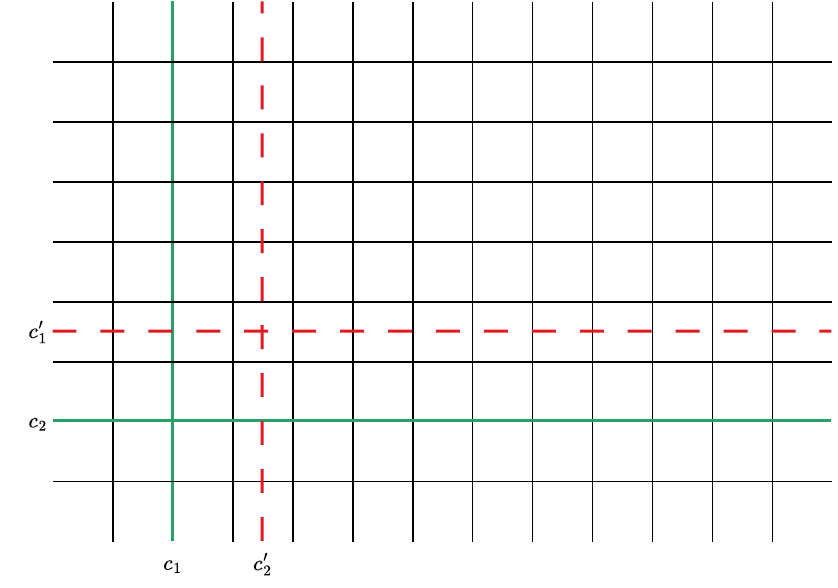}
    \caption{The two non-contractible cycles of the torus. Solid green lines are the primal-lattice cycles $c_1$, $c_2$, defining $W^Z_{c_1}$, $W^Z_{c_2}$; dashed red lines are the dual-lattice cycles $c'_1$, $c'_2$, defining $W^X_{c'_1}$, $W^X_{c'_2}$}.
    \label{fig:c1-c2}
\end{figure}

In this paper, we focus on toric code on a torus.  
This results in a 4-fold ground-state degeneracy which can be fixed by finding two extra independent and commuting operators. A natural choice is to pick one of the four pairs below: 
\begin{equation}\label{eq:pairs}
    (W^Z_{c_1}, W^Z_{c_2}),
\, (-W^Z_{c_1}, W^Z_{c_2}),\, (W^Z_{c_1}, -W^Z_{c_2}), \, (-W^Z_{c_1}, -W^Z_{c_2}), 
    \end{equation} 
where $W^Z_{c_1}$ and $W^Z_{c_2}$ are products of $Z$s along the two non-contractible cycles $c_1$ and $c_2$ of the torus as shown in Fig. \ref{fig:c1-c2}. As an error-correcting code, the interpretation is that two qubits are encoded into the four ground states of the toric code. 

\begin{figure}
    \centering
    \includegraphics[width=\linewidth]{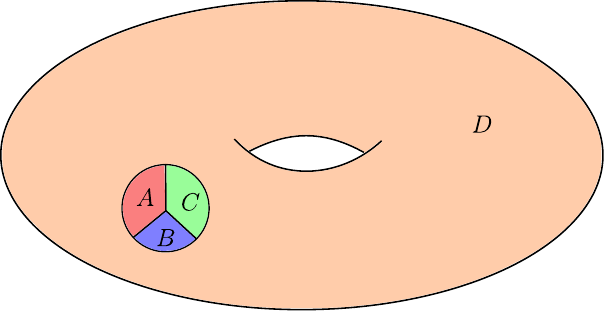}
    \caption{Kitaev--Preskill disk partition \cite{Kitaev:2005dm} for subsystem $A$, $B$, and $C$.}
    \label{fig:partition-kp}
\end{figure}

To calculate the $n = 4$  GM, we need to specify how to partition the edges of the toric lattice into four parties $A$, $B$, $C$, and $D$. We follow the same prescription as \cite{Kitaev:2005dm} and partition the torus as shown in Fig. \ref{fig:partition-kp},  which we call a Kitaev--Preskill partition.
For a Kitaev--Preskill partition, the following two conditions must be satisfied: (i) $A \cup B \cup C$ has the topology of a disk. In particular, none of $A$, $B$, or $C$ contains a non-contractible cycle of the torus. (ii) There are exactly  \textit{four junction points} at which three regions meet,  and no points at which all four regions meet. On the lattice,  junction points can be either lattice vertices, or centers of plaquettes. We denote the number of junction points on the lattice vertices as $n_{\bullet}$, and that at plaquette centers as  $n_{\circ}$, thus $n_{\bullet}+n_{\circ}=4$ for Kitaev--Preskill partitions. We present evidence that GM distinguishes two cases: $n_{\bullet} \neq n_{\circ}$ and $n_{\bullet} = n_{\circ}$, and 
that deformations of the regions $A$, $B$, $C$,
or $D$ that preserve the Kitaev--Preskill conditions and remain within each case 
do not affect the numerical value of the GM.

Moreover, we also present evidence that increasing the lattice size $L$ also does not affect the GM. There is a simple heuristic explanation of these facts if the state is stabilizer. From the construction of the GM, the GM is non-zero only when the stabilizer state has stabilizer group elements that act non-locally on all four parties $A$, $B$, $C$, and $D$. One way to produce such elements is by taking products of group elements with support on $A \cup B$, $B \cup C$, $C \cup D$, etc. However, such a product element does not generate \emph{genuine} multipartite entanglement, and is expected to cancel in the GM. The only elements that generate genuine multipartite entanglement are ones that cannot be written as a product of lower-partite stabilizer generators.

In the toric code, the stabilizer elements are topologically trivial closed loops of $Z$s on the lattice, {\it i.e.}, $B_p$, or topologically trivial closed loops of $X$s on the dual lattice, {\it i.e.}, $A_v$. These are supplemented with the two non-contractible cycles $W_{c_1}^{Z}$ and $W_{c_2}^{Z}$. Given any four-partite decomposition of the edges of the toric lattice, only an $\mathcal{O}(1)$ number of stabilizer elements can have non-trivial support on all four regions $A$, $B$, $C$, and $D$, while not being expressible as products of elements supported on at most three regions. Such elements must be associated with the non-contractible cycles and, crucially, their number does not grow with the system size $L$. This heuristic argument suggests that any genuine multipartite measure of entanglement doesn't scale with $L$, unlike the entanglement entropy which satisfies the famous $\alpha L - \gamma$ formula.

These arguments can be made more concrete if we work with the genuine multipartite entanglement measure introduced in \cite{Fattal:2004frh} for stabilizer states. For a $\mathtt{q}$-partite stabilizer state, their measure is the number of such non-local stabilizer group elements that cannot be built out of group elements with smaller support. Suppose $G$ is the stabilizer group of a four-partite stabilizer state. The measure introduced in \cite{Fattal:2004frh} is 
\begin{equation}
    p = \log_2 \left(\frac{|G|}{|G_{ABC}\cdot G_{ABD} \cdot G_{ACD} \cdot G_{BCD}|}\right), 
\end{equation}
where $|G|$ denotes the order of the group $G$ and $G_{ABC}\cdot G_{ABD} \cdot G_{ACD} \cdot G_{BCD}$ is the group of all stabilizer elements which are products of stabilizer elements with support over three or fewer number of parties. As argued in \cite{Bravyi:2005ztn}, $p$ is bounded above by an $\mathcal{O}(1)$ number that does not scale with the lattice size $L$ for the toric code.

In the next section, we study the GM for the stabilizer ground state of the toric code defined on an $L \times L$ lattice for small values of $L \le 7$. As discussed above, the stabilizer ground state is stabilized by all the star and plaquette operators, along with the $W_{c_1}^{Z}$ and $W_{c_2}^{Z}$ operators with support over the non-contractible loops of the torus.  The other three stabilizer ground states can be obtained by acting with $W_{c_1'}^{X}$, $W_{c_2'}^{X}$, and $W_{c_1'}^{X}W_{c_2'}^{X}$ respectively, where $W_{c_1'}^{X}$ and $W_{c_2'}^{X}$ are products of $X$'s along the two non-contractible loops in the dual lattice as shown with the dashed red lines in Fig. \ref{fig:c1-c2}. As these are a local product of Pauli operators, the GM cannot distinguish the four stabilizer ground states.

\section{GM for the Stabilizer Ground-State}
\label{StabilizerGM}

We study the behavior of the four-partite GM
for the stabilizer ground states. As summarized in Sec.~\ref{Intro}, a distinctive feature is
a threshold at the R\'enyi index $n=4$.

For $n=2$ \cite{Iizuka:2025pqq} and $n=3$, we find that the GM reduces to the
R\'enyi tripartite information:
\begin{equation}
\begin{aligned}
\GM[\mathtt{q}=4]_{n=2}
&=
-\left(a-\frac{1}{12}\right) I_3,
\\
\GM[\mathtt{q}=4]_{n=3}
&=
-\left(a-\frac{1}{9}\right) I_3.
\end{aligned}
\label{n23GMI3}
\end{equation}
In particular, for $a=1/3$,
\begin{align}
\GM[\mathtt{q}=4]_{n=2}
&=
-\frac14 I_3,
&
\GM[\mathtt{q}=4]_{n=3}
&=
-\frac29 I_3.
\end{align}
Thus, for the Kitaev--Preskill partition, the low-$n$ GM
is completely fixed by the same total-quantum-dimension
data that determines the topological entanglement entropy. We confirm these relations numerically for
$10^5$ randomly sampled stabilizer states\footnote{In each of the $10^5$ runs, we first pick a random integer between $4$ and $16$ for the number of qubits of the stabilizer state and randomly partition these qubits into four parties. Then we pick another random number between $0$ and $1$ and consider the Erd\H{o}s--R\'enyi random graph where each edge independently appears with this probability. The stabilizer state is the associated graph state.}.

However, at $n=4$, this reduction no longer holds. Indeed, the ratio
${\GM[\mathtt{q}=4]_{n=4}}/{I_3}$
is not constant across the four-partitions considered here. In this section, we present numerical evidence that, 
for Kitaev--Preskill
partitions of the toric-code stabilizer ground states at $a=1/3$,
\begin{equation}
\GM[\mathtt{q}=4]_{n=4}
=
\begin{cases}
\frac{3}{16}\log (2), & n_\bullet\neq n_\circ, \quad n_\bullet+n_\circ=4 \\[3pt]
\frac{11}{48}\log (2), & n_\bullet=n_\circ=2.
\end{cases}
\label{eq:KPn4q4relation}
\end{equation}
Within each case, the GM is invariant under deformations that preserve the Kitaev--Preskill conditions defined in the previous section, and is independent of the lattice size.  We call these deformations \emph{small} and use ``topological invariance'' in this restricted sense. We confirmed Eq.~\eqref{eq:KPn4q4relation} by a systematic scan of $\sim 3\times10^4$ single-edge deformations of the Kitaev--Preskill partitions. 
Note that the GM at $n=2$ and $n=3$ is also topological, since it is proportional to $I_3$, though it does not distinguish the two cases of Kitaev--Preskill partitions.

We illustrate Eq.~\eqref{eq:KPn4q4relation} through several explicit examples, and we show that the ratio ${\GM[\mathtt{q}=4]_{n=4}}/{I_3}$ takes different values, thereby showing that the ${\GM[\mathtt{q}=4]_{n=4}}$ does not collapse to $I_3$. We also compare our result with the general result of \cite{DelZotto:2026fpw} where they argue that any genuine multipartite signal is given by the partition function of the underlying TQFT up to $\mathcal{O}(e^{-cL})$ corrections. For the toric code, our analysis shows that no such corrections are needed.

We then compare the toric code with the color code \cite{Bombin:2006sj, Bombin:2006rc} on a honeycomb lattice. As shown in \cite{Bombin:2011qp}, the underlying TQFT of the color code is equivalent to two copies of the toric code TQFT. The $n = 4$ GM in the color code is also a topological invariant, and is twice that of the $n = 4$ GM in the toric code. This provides a consistency check to our claim that the $n = 4$ GM is a topological invariant of topological stabilizer codes.

\subsection{Topological Invariance}

\begin{figure*}
\centering
\begin{minipage}[t]{0.24\textwidth}\centering
\includegraphics[width=\linewidth]{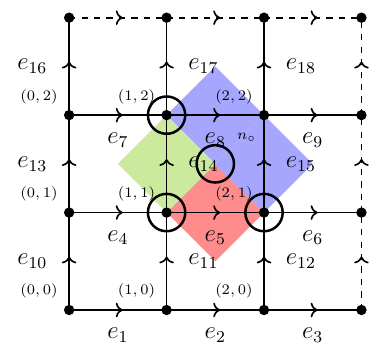}
\end{minipage}\hfill
\begin{minipage}[t]{0.24\textwidth}\centering
\includegraphics[width=\linewidth]{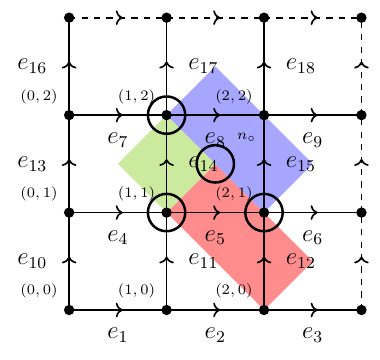}
\end{minipage}\hfill
\begin{minipage}[t]{0.24\textwidth}\centering
\includegraphics[width=\linewidth]{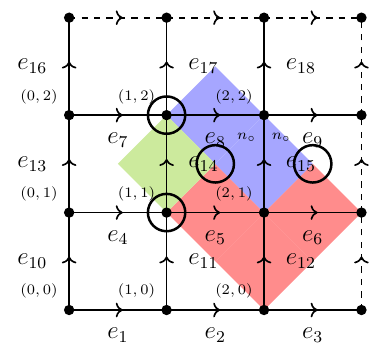}
\end{minipage}\hfill
\begin{minipage}[t]{0.24\textwidth}\centering
\includegraphics[width=\linewidth]{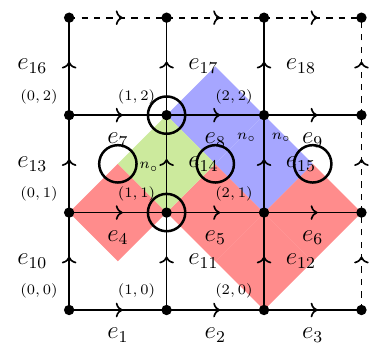}
\end{minipage}
\caption{Starting from the Kitaev--Preskill partition on the $3\times3$ toric lattice in the first panel, we successively add $e_{12}$, $e_6$, and $e_4$ to region $A$ to obtain the second, third, and fourth panels, respectively. Black circles mark junction points where three regions meet. One can confirm that Eq.~\eqref{eq:KPn4q4relation} holds for the first three panels. For the fourth panel, $A$ contains a non-contractible loop, so the partition is no longer Kitaev--Preskill; nevertheless, the GM remains unchanged from the third panel by accident.}
\label{fig:3x3-kp-deformations}
\end{figure*}

Let's consider the Kitaev--Preskill partition of the $3\times3$ lattice shown in the first  panel of Fig.~\ref{fig:3x3-kp-deformations}, where $n_\bullet=3$, $n_\circ=1$. For the stabilizer ground states of the toric code, we  find that for $a=1/3$,
\begin{equation}
    \text{GM}_{n = 4}^{(\mathtt{q} = 4)}(A:B:C:D)|_{3\times 3} = \frac{3}{16} \log(2),
\end{equation}
which agrees with Eq.~\eqref{eq:KPn4q4relation}. Let us extend region $A$ by $e_{12}$ (second panel), where $n_\bullet=3$, $n_\circ=1$. This is a small deformation and one can confirm that the GM is unchanged. Extending $A$ further by $e_6$ (third panel) is \emph{not} a small deformation, since it is still a Kitaev--Preskill partition, but  $n_{\bullet} = n_{\circ} = 2$. For this partition, we find 
\begin{equation}
    \text{GM}_{n = 4}^{(\mathtt{q} = 4)}(A:B:C:D)|_{3\times 3} = \frac{11}{48} \log(2),
\end{equation}
which agrees with Eq.~\eqref{eq:KPn4q4relation}. 
Extending $A$ once more by $e_4$ (fourth panel) gives rise to a partition that is no more a Kitaev--Preskill partition as $A$ now contains a noncontractible cycle. We now also have $n_{\bullet} + n_{\circ} = 5$ instead of $4$. Nevertheless, the GM turns out to be $11/48\log(2)$, but this is just an accident, not a consequence of anything established here. We could not identify any rule for partitions that are not Kitaev--Preskill type. 

\begin{figure}
    \centering
    \includegraphics[width=\linewidth]{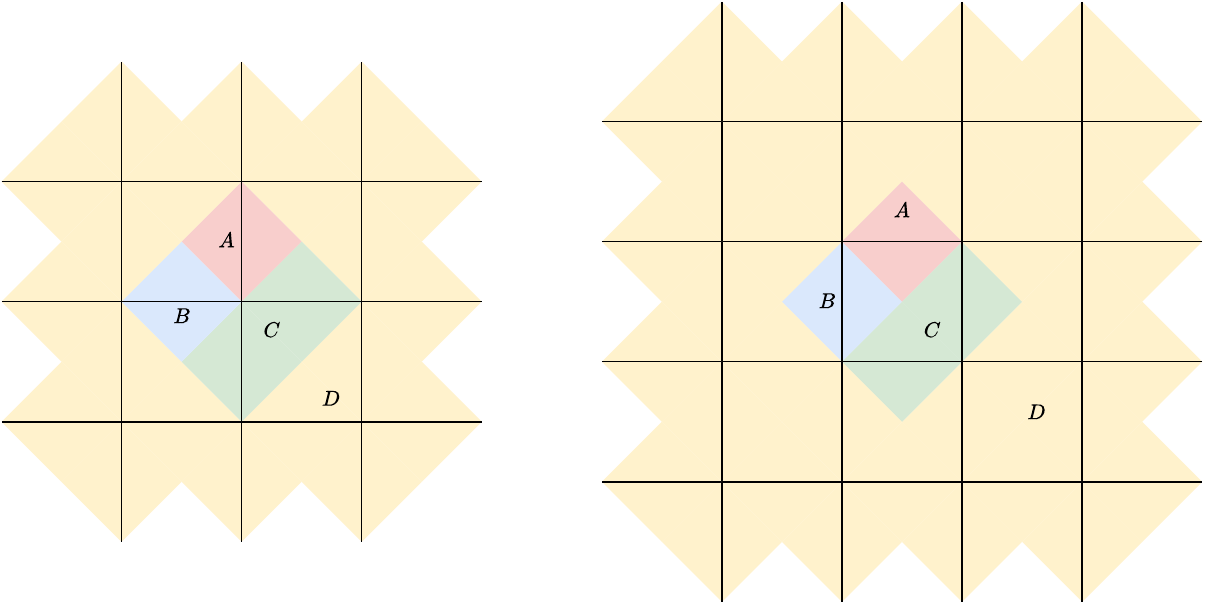}
    \caption{In both cases, the GM is ${3}/{16}\log(2)$, demonstrating that changing the lattice size $L$ does not affect the GM.}
    \label{fig:3x3-4x4-partition-kp}
\end{figure}

Let's now consider bigger lattices. The same Kitaev--Preskill partition for a $3\times 3$ and $4 \times 4$ lattice is shown in Fig.~\ref{fig:3x3-4x4-partition-kp}. In both cases, the GM continues to be $3/16 \log(2)$, demonstrating that changing the lattice size has no effect on the GM for this simple Kitaev--Preskill partition. We have confirmed that the topological invariance of the GM within the two cases of Kitaev--Preskill partitions continues to hold for $L \times L$ lattices up to $L \le 7$.

Our analysis also demonstrates that there are no $\mathcal{O}(e^{-cL})$ corrections to the GM. This appears to be a novel feature of the toric code. For more general string-net Hamiltonians, the GM of the ground state is expected to pick up some $L$-dependent corrections which vanish in the $L \to \infty$ limit to extract some topological invariants. In the toric code, we directly get the topological invariants without needing the $L \to \infty$ limit. 

\begin{figure}
    \centering
    \includegraphics[width=0.5\linewidth]{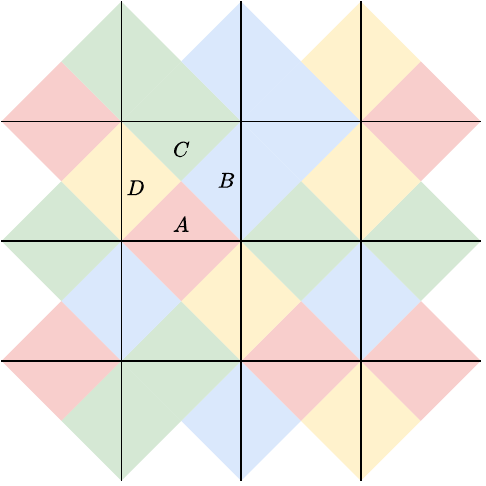}
    \caption{An alternative four-partition of the $3 \times 3$ toric-code lattice, in which $A$, $B$, $C$, $D$ are arranged differently from Fig.~\ref{fig:3x3-4x4-partition-kp}. For this partition, $\GM_{n=4}=7/16 \log(2)$ and $I_3=-3 \log(2)$, so the ratio $\GM_{n=4}/I_3$ differs from the Kitaev--Preskill partition.}
    \label{fig:3x3-partition-violation}
\end{figure}

Let's now show that the $n = 4$ GM does not collapse to the tripartite information $I_3$. For the Kitaev--Preskill partition of the $3 \times 3$ toric lattice in Fig. \ref{fig:3x3-4x4-partition-kp}, the values of $I_3$ and the $n = 4$ GM for $a = 1/3$ are: 
\begin{equation}
\begin{split}
    & \text{GM}_{n=4}^{(\mathtt{q} = 4)}(A:B:C:D)|_{3 \times 3} = \frac{3}{16} \log(2), \\ & I_3 = - \log(2).
\end{split}
\end{equation}
Now, consider the partition of the $3 \times 3$ lattice as shown in Fig. \ref{fig:3x3-partition-violation}. The numerical values of the $n = 4$ GM, for $a = 1/3$, and that of $I_3$ are 
\begin{equation}
    \begin{split}
        & \text{GM}_{n=4}^{(\mathtt{q} = 4)}(A:B:C:D)|_{3\times 3} = \frac{7}{16} \log(2), \\
        & I_3 = -3 \log(2).
    \end{split}
\end{equation}
The ratio for this partition is different from the Kitaev--Preskill partition. This, therefore, demonstrates that the $n = 4$ GM does not collapse to the tripartite information $I_3$; unlike the $n = 2$ or the $n = 3$ GM. 

\subsection{Comparison to the color code} 

In the previous subsection, we argued that the $n = 4$ GM is a topological invariant for stabilizer ground states of the toric code within the deformation class defined there, and provided an explicit example where the GM does not collapse to the tripartite information $I_3$. In this subsection, we compare the toric code to the 2D color code \cite{Bombin:2006sj, Bombin:2006rc, Bombin:2006cd}. The color code, similar to the toric code, is a topological stabilizer error-correcting code whose underlying TQFT is two copies of the toric-code TQFT as shown in \cite{Bombin:2011qp}. Therefore, the $n = 4$ GM in the color code, if it is a topological invariant, must be twice that of the toric code. We numerically verify this for both the $n = 4$ GM and the tripartite information $I_3$ for stabilizer states in the color code.

\begin{figure}
    \includegraphics[width=\linewidth]{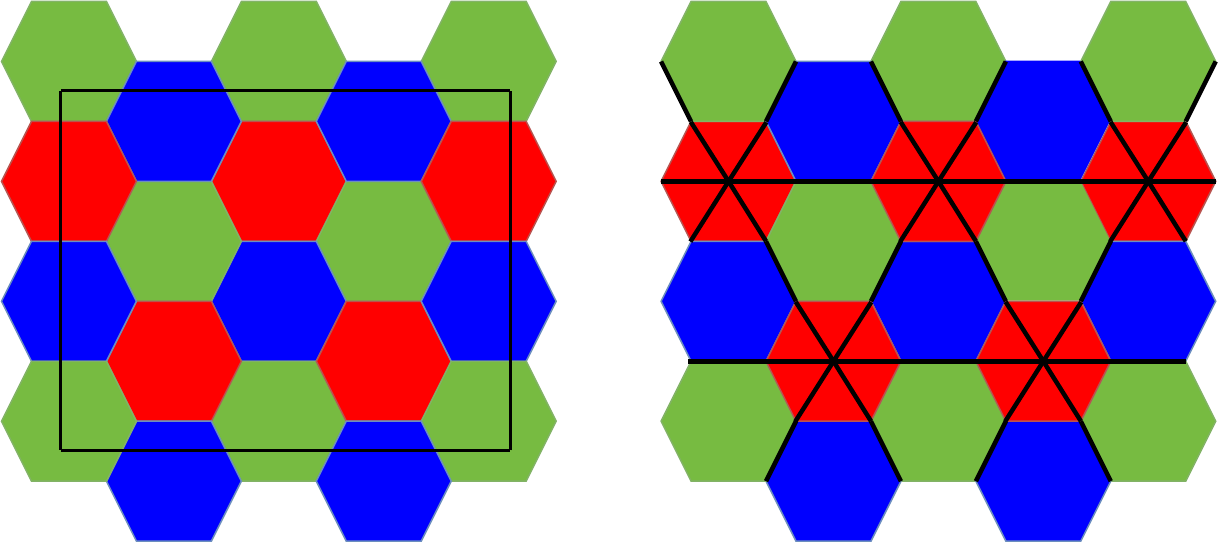}
    \caption{(Left) Honeycomb lattice for the 2d color code. Each face carries three colors such that no two adjacent faces have the same color. (Right) The red shrunk lattice obtained by shrinking each red face to a point and joining them together by edges. Shrunk lattices of the honeycomb lattice are triangular.}
    \label{fig:torus-honeycomb}
\end{figure}

To be concrete, let's consider a honeycomb lattice on the surface of a torus such that each hexagonal face carries one of three colors: red, blue, or green, and the lattice is such that any two adjacent hexagons have different colors. An example is shown in Fig. \ref{fig:torus-honeycomb}. The rectangle denotes periodic boundary conditions of the torus. The qubits live on the vertices of the lattice, and the following stabilizer generators are associated to every hexagonal face of the lattice: 
\begin{equation}
    X_f = \bigotimes_{v \in f} X_v, \quad Z_f = \bigotimes_{v \in f} Z_v. 
\end{equation}
On the honeycomb lattice, these generators commute with each other. 

On a genus $g$ surface, the color code encodes $4g$ logical qubits while the toric code encodes $2g$ logical qubits. The ground-state degeneracy, therefore, needs to be fixed by choosing $4$ additional stabilizer generators that commute with every $X_f$ and $Z_f$. The canonical way to choose these operators is to pick non-contractible loops on the so called \emph{shrunk lattice}. There is a shrunk lattice we can build from each of the three colors. The shrunk lattice is obtained by shrinking, say, the red faces to a vertex and joining these vertices by edges. The shrunk lattice is triangular as shown in the right panel of Fig. \ref{fig:torus-honeycomb}. Let $c_1$ and $c_2$ be the two non-contractible cycles on the shrunk lattice on the torus. We can associate $W^Z_{c_1}$, $W^Z_{c_2}$ to each of the three shrunk lattices. There are only 4 independent such generators, and supplementing the $X_f$ and $Z_f$ with them lifts the degeneracy. Other choices can be obtained by acting with a local product of Pauli $X$ operators; exactly like the toric code. 

\begin{figure}
    \centering
    \includegraphics[width=0.75\linewidth]{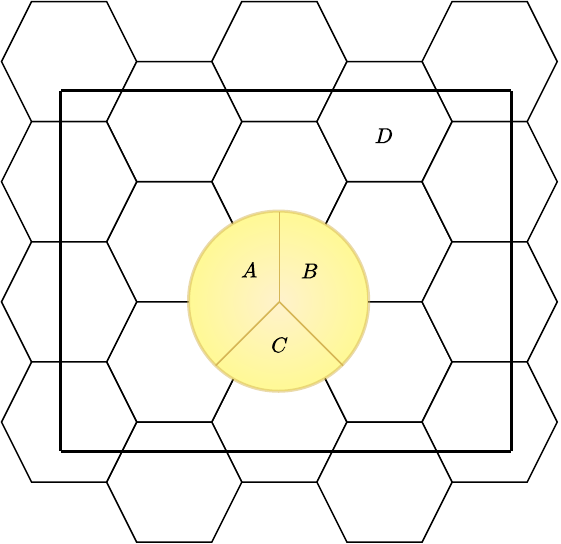}
    \caption{Kitaev--Preskill partition of the honeycomb lattice to calculate the $n = 4$ GM and $I_3$.}
    \label{fig:honeycomb-kp-partition}
\end{figure}

As with the toric code, we consider a Kitaev--Preskill partition of the vertices of the honeycomb lattice as shown in Fig. \ref{fig:honeycomb-kp-partition}. The numerical value of the $n = 4$ GM for $a = 1/3$, and that of $I_3$, is precisely twice that of the toric code: 
\begin{equation}
\begin{split}
    & \text{GM}_{n=4}^{(\mathtt{q}=4)}(A:B:C:D) = \frac{3}{8} \log(2), \\ 
    & I_3 = - 2 \log(2).
\end{split}
\end{equation}
Similar to the toric code, these numbers are robust to small deformations, and independent of the lattice size. Moreover, they are consistent with the expectation that the underlying TQFT of the color code on the honeycomb lattice is two copies of the toric code TQFT \cite{Bombin:2011qp}.

\section{Beyond the Stabilizer Ground-State Sector}
\label{sec:beyond-stabilizer}

The stabilizer analysis provides a controlled starting point for
understanding genuine multi-entropy in the toric code. However, the
physical motivation of the present work is broader. We would like to
distinguish which features of genuine multi-entropy are special to the
stabilizer description, which features are associated with the
topological ground-state sector more generally, and how these structures
are modified by local excitations.

\subsection{The $2\times 2$ lattice setup:\\ General ground-state superpositions}
\label{subsec:general-ground-states}

To address these questions, we study general ground-state
superpositions on the $2\times2$ toric-code lattice. We first specify a
convenient stabilizer basis of the four-dimensional ground-state
subspace and then consider general linear combinations of the basis
states.

In order to study GM for generic ground-state superpositions, we evaluate the relevant replica contractions numerically. We focus on the toric code on a $2\times2$ square lattice with periodic boundary conditions. This is the smallest and simplest toroidal lattice that retains the four-dimensional topological ground-state manifold and admits a nontrivial four-partite partition of the edge degrees of freedom. It therefore provides a minimal setting in which stabilizer ground states, non-stabilizer superpositions within the same ground-state manifold, and locally excited states can all be compared explicitly. The physical qubits live on the edges of the lattice, and we label the eight independent edge degrees of freedom as follows.
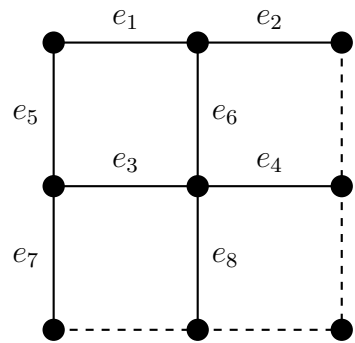
\begin{figure}[h]
\centering
\begin{tikzpicture}[
    scale=1.9,
    every node/.style={font=\large},
    edge/.style={thick},
    copyedge/.style={thick,dashed}
]

\foreach \x in {0,1,2} {
  \foreach \y in {0,1,2} {
    \fill (\x,-\y) circle (2.2pt);
  }
}

\draw[edge] (0,0) -- (1,0);
\node[above=2pt] at (0.5,0) {$e_1$};

\draw[edge] (1,0) -- (2,0);
\node[above=2pt] at (1.5,0) {$e_2$};

\draw[edge] (0,-1) -- (1,-1);
\node[above=2pt] at (0.5,-1) {$e_3$};

\draw[edge] (1,-1) -- (2,-1);
\node[above=2pt] at (1.5,-1) {$e_4$};

\draw[copyedge] (0,-2) -- (1,-2);
\draw[copyedge] (1,-2) -- (2,-2);

\draw[edge] (0,0) -- (0,-1);
\node[left=2pt] at (0,-0.5) {$e_5$};

\draw[edge] (1,0) -- (1,-1);
\node[right=2pt] at (1,-0.5) {$e_6$};

\draw[edge] (0,-1) -- (0,-2);
\node[left=2pt] at (0,-1.5) {$e_7$};

\draw[edge] (1,-1) -- (1,-2);
\node[right=2pt] at (1,-1.5) {$e_8$};

\draw[copyedge] (2,0) -- (2,-1);
\draw[copyedge] (2,-1) -- (2,-2);

\end{tikzpicture}

\caption{
The $2 \times 2$ toric-code lattice with periodic boundary conditions.
The physical qubits live on the eight independent edges labeled by
$e_1,\ldots,e_8$. Solid edges denote independent degrees of freedom,
while dashed edges denote their periodic copies.
}
\label{Fig:2by2lattice}
\end{figure}

Here the right boundary is identified with the left boundary, and the
bottom boundary is identified with the top boundary. Thus the repeated
edges in the above drawing are not additional degrees of freedom. 
The Hilbert space has dimension $2^8=256$. 
Appendix~\ref{Apptoric2by2} gives the star and plaquette operators
explicitly in terms of $X_i$ and $Z_i$, where $X_i$ and $Z_i$ denote
the Pauli $X$ and $Z$ operators acting on the qubit associated
with the edge $e_i$, respectively.

To specify a convenient basis of the four-dimensional ground-state subspace, we choose the two non-contractible $Z$-loop operators
\begin{equation}
W^Z_x=Z_1 Z_2 \sim Z_3 Z_4,
\qquad
W^Z_y=Z_5Z_7 \sim Z_6 Z_8.\label{eq:loop_operators}
\end{equation}
where the equivalences hold within the ground-state subspace. 
They commute with the toric-code Hamiltonian and with each other. We
therefore label the ground states by their eigenvalues,
\begin{equation}
W^Z_x|\psi_{\alpha\beta}\rangle
=
(-1)^\alpha|\psi_{\alpha\beta}\rangle \,,
\quad
W^Z_y|\psi_{\alpha\beta}\rangle
=
(-1)^\beta|\psi_{\alpha\beta}\rangle\,,
\label{eq:defofalphabeta}
\end{equation}
where
$\alpha,\beta=0,1$. 

The corresponding non-contractible $X$-loop operators may be chosen as
\begin{equation}
W^X_x=X_1X_3 \sim X_2X_4\,,
\quad
W^X_y=X_5X_6 \sim X_7X_8,
\end{equation}
again, the equivalences hold within the ground-state subspace. Unlike the
$Z$-loops, these $X$-loop operators change the topological sector:
\begin{equation}
W^X_x|\psi_{\alpha\beta}\rangle
=
|\psi_{\alpha+1,\beta}\rangle \,,
\quad
W^X_y|\psi_{\alpha\beta}\rangle
=
|\psi_{\alpha,\beta+1}\rangle,
\end{equation}
where the additions are modulo $2$. Thus, starting from
$|\psi_{00}\rangle$, the remaining basis states are generated as
\begin{equation}
|\psi_{\alpha\beta}\rangle
=
(W^X_x)^\alpha
(W^X_y)^\beta
|\psi_{00}\rangle.
\end{equation}
Explicit computational-basis expressions for
$\ket{\psi_{\alpha\beta}}$ on the $2\times2$ toric-code lattice are
given in Appendix~\ref{Apptoric2by2}.

It follows that the four ground states are related by local unitary transformations with respect to any partition of the edge degrees of freedom, since $W^X_x$ and $W^X_y$ are products of single-edge Pauli operators. Since genuine multi-entropy is invariant under local unitaries, its value is identical for all four stabilizer ground states:
\begin{equation}
\GM[4]_n\left(|\psi_{\alpha\beta}\rangle\right)
=
\GM[4]_n\left(|\psi_{00}\rangle\right),
\qquad
\alpha,\beta=0,1.
\end{equation}
The four states $|\psi_{\alpha\beta}\rangle$ form a stabilizer basis of
the ground-state subspace. In particular, $|\psi_{00}\rangle$ is
stabilized by the star and plaquette operators together with the two
non-contractible $Z$-loop operators. 
By contrast, a general normalized ground state can be written as
\begin{equation}
|\Psi\rangle
=
\sum_{\alpha,\beta=0,1}
c_{\alpha\beta}|\psi_{\alpha\beta}\rangle \,, \quad \mbox{where} \,\,
\sum_{\alpha,\beta=0,1}|c_{\alpha\beta}|^2=1.
\label{eq:general-ground-state}
\end{equation}
For generic coefficients $c_{\alpha\beta}$, the state $|\Psi\rangle$
need not be a stabilizer state. This distinction allows us to compare
the behavior of genuine multi-entropy in the stabilizer sector with its
behavior for non-stabilizer states within the same topological
ground-state manifold.

Our choice of four-partite partition is the ``checkerboard'' partition, 
\begin{equation}
\begin{aligned}
A&=\{e_1,e_5\},\qquad
B=\{e_2,e_6\}, \\
C&=\{e_3,e_7\},\qquad
D=\{e_4,e_8\}.
\label{checkerboard}
\end{aligned}
\end{equation}
shown in Fig.~\ref{Fig:2by2lattice}. 
For this partition, we numerically evaluate the GM for the following four families:

\subsubsection{A one-parameter non-stabilizer family}

First, we consider the simplest real one-parameter family
\begin{equation}
|\Psi_{10}(\theta)\rangle
=
\cos\theta\,|\psi_{00}\rangle
+
\sin\theta\,|\psi_{10}\rangle ,
\quad
0\leq \theta \leq \frac{\pi}{2}.
\label{eq:Psi10_theta}
\end{equation}
We compute $\GM[\mathtt{q}=4]$ with $a=1/3$, together with the tripartite information $I_{3,n}$, for several discrete choices of $\theta$ and for $n=2,3$. Our purpose is to examine whether the relation between $\GM[\mathtt{q}=4]$ and $I_{3,n}$, which holds for stabilizer states, continues to hold away from the stabilizer points.

\begin{figure}[t]
\centering

\includegraphics[width=0.95\columnwidth]{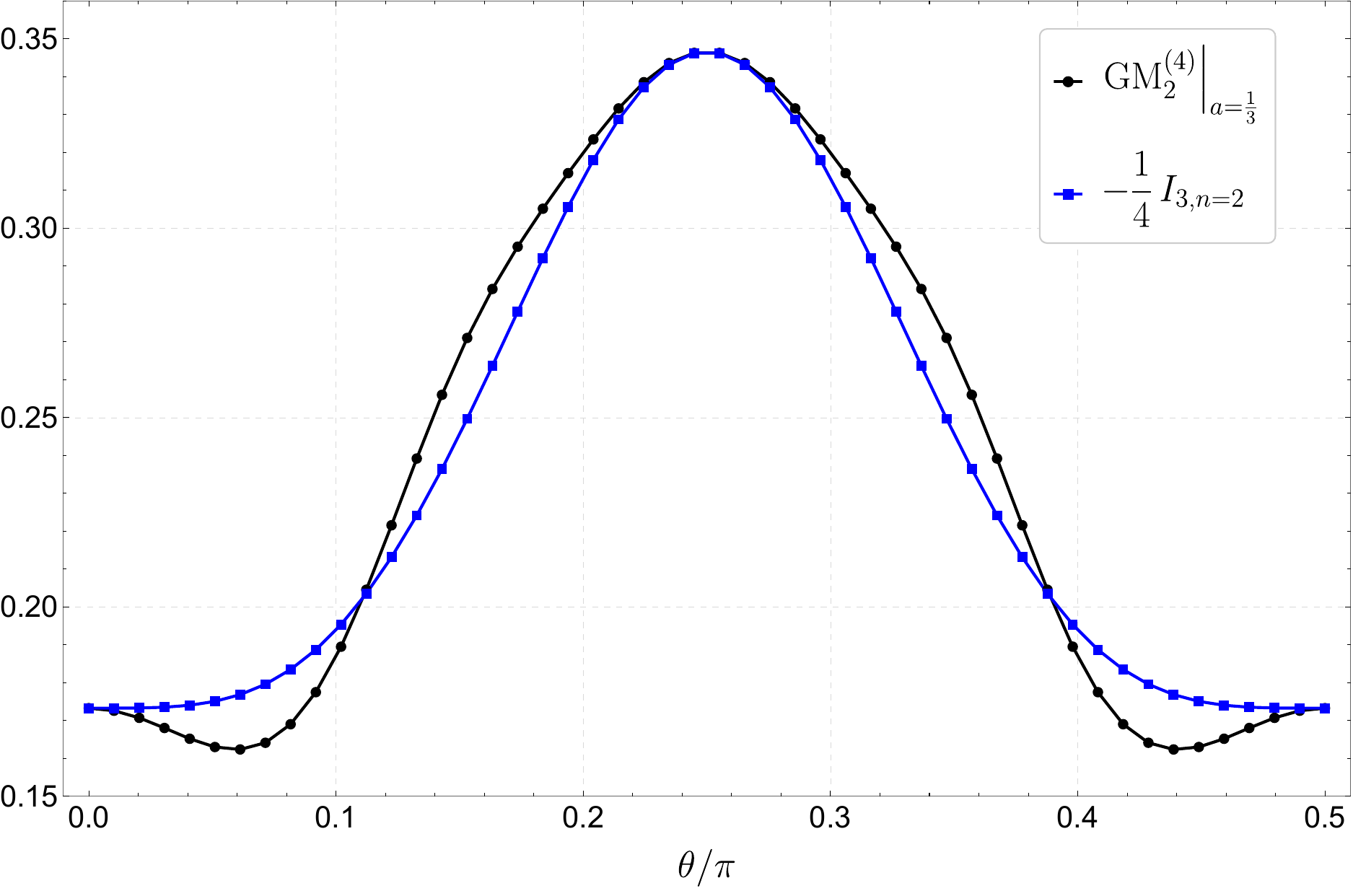}

\vspace{0.8em}

\includegraphics[width=0.95\columnwidth]{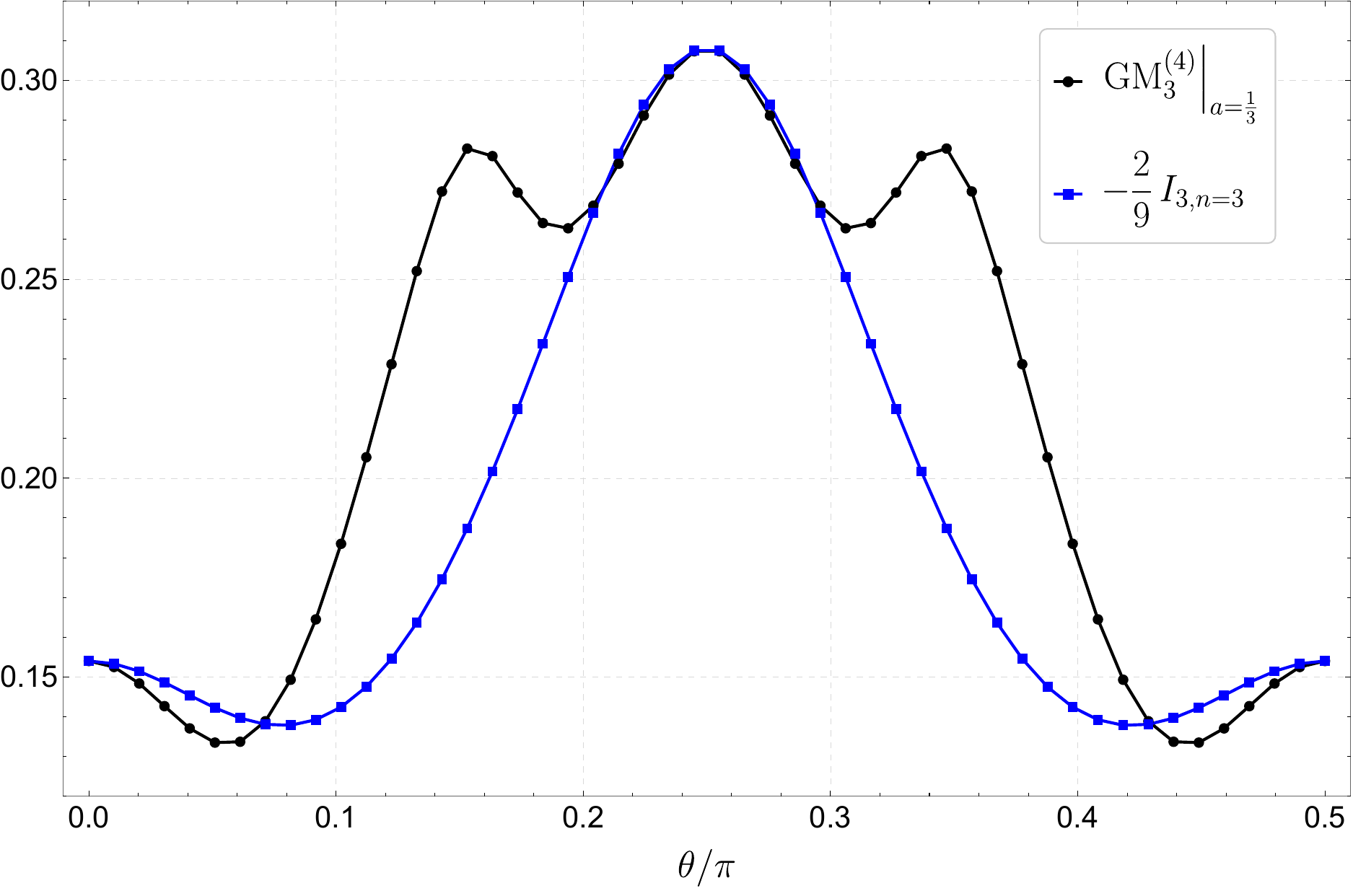}

\caption{
$\GM[\mathtt{q}=4]_n$ at $a=1/3$ for
$|\Psi_{10}(\theta)\rangle$.
Upper panel: $n=2$, together with $-I_{3,n=2}/4$.
The relation \eqref{eq:GMn2},
$\GM[\mathtt{q}=4]_{n=2}=-I_{3,n=2}/4$,
holds at the stabilizer points
$\theta=0,\pi/4,\pi/2$, while it is generically violated away from
these points.
Lower panel: $n=3$, together with $-2I_{3,n=3}/9$.
The relation \eqref{eq:GMn3},
$\GM[\mathtt{q}=4]_{n=3}=-2I_{3,n=3}/9$,
again holds at the stabilizer points
$\theta=0,\pi/4,\pi/2$.
For generic non-stabilizer values of $\theta$, the two quantities do
not agree.
}
\label{fig:Psi10}
\end{figure}

The result for $n=2$ and $n=3$ with $a=1/3$ choice is shown in Fig.~\ref{fig:Psi10}.  The upper panel shows $\GM[\mathtt{q}=4]_{n=2}$ and $-I_{3,n=2}/4$ as functions of $\theta$. The lower panel shows $\GM[\mathtt{q}=4]_{n=3}$ and $-2I_{3,n=3}/9$. 
At the stabilizer points $\theta=0,\pi/4,\pi/2$, the relations
\eqref{eq:GMn2} and \eqref{eq:GMn3} reduce, for $a=1/3$, to
\begin{align}
\left.\GM[\mathtt{q}=4]_{n} \right|_{a=1/3}
&=
\lambda_n I_{3,n} \quad \mbox{($n=2, 3$)}\,,  \\
\lambda_{n=2} = -\frac{1}{4} \,, &\quad \lambda_{n=3} = -\frac{2}{9} 
\end{align}
For generic values of $\theta$, however, GM and the corresponding
multiple of $I_{3,n}$ do not agree.
Note that there are isolated values of $\theta$ at which GM and the corresponding
multiple of $I_{3,n}$ happen to coincide. However, these accidental intersections occur at different values of $\theta$ for $n=2$ and $n=3$, and therefore do not indicate a universal relation. One can also confirm in all the $\theta$, 
\begin{equation}
    - I_{3,n} > \left.\GM[\mathtt{q}=4]_{n} \right|_{a=1/3} > 0  
\end{equation}
is satisfied for both $n=2$ and $n=3$. 

These results show that the proportionality relation between $\GM[\mathtt{q}=4]_n$ and $I_{3,n}$ found for stabilizer states does not extend to a general linear combination of toric-code ground states. In particular, it is not an identity throughout the ground-state Hilbert space. Rather, it is a special property of the stabilizer loci within this one-parameter family. These show that $\GM$ and $I_{3,n}$ are physically different quantities. 

\subsubsection{Another one-parameter non-stabilizer family}

\begin{figure}[h]
\centering

\includegraphics[width=0.95\columnwidth]{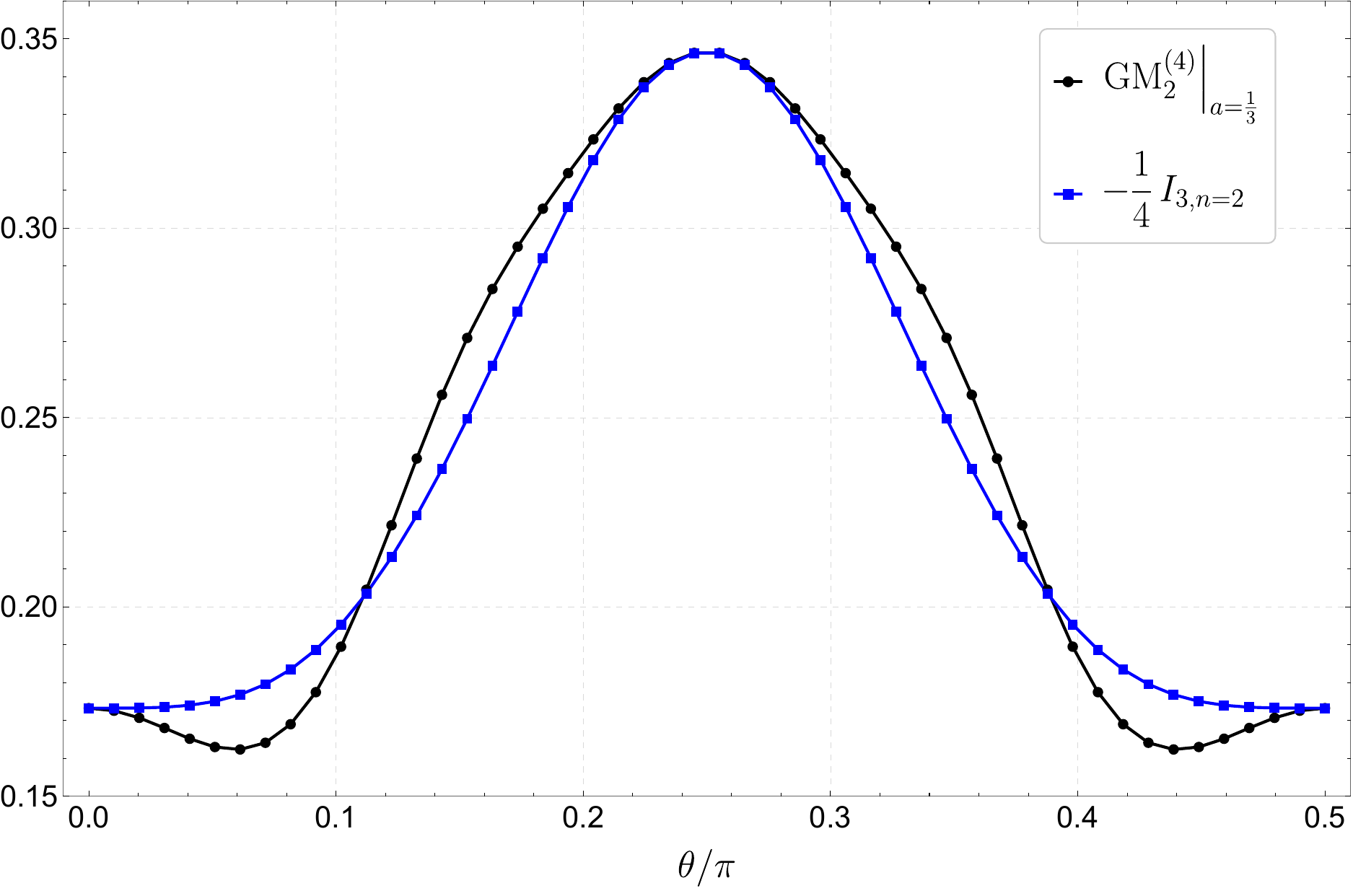}

\vspace{0.5cm}

\includegraphics[width=0.95\columnwidth]{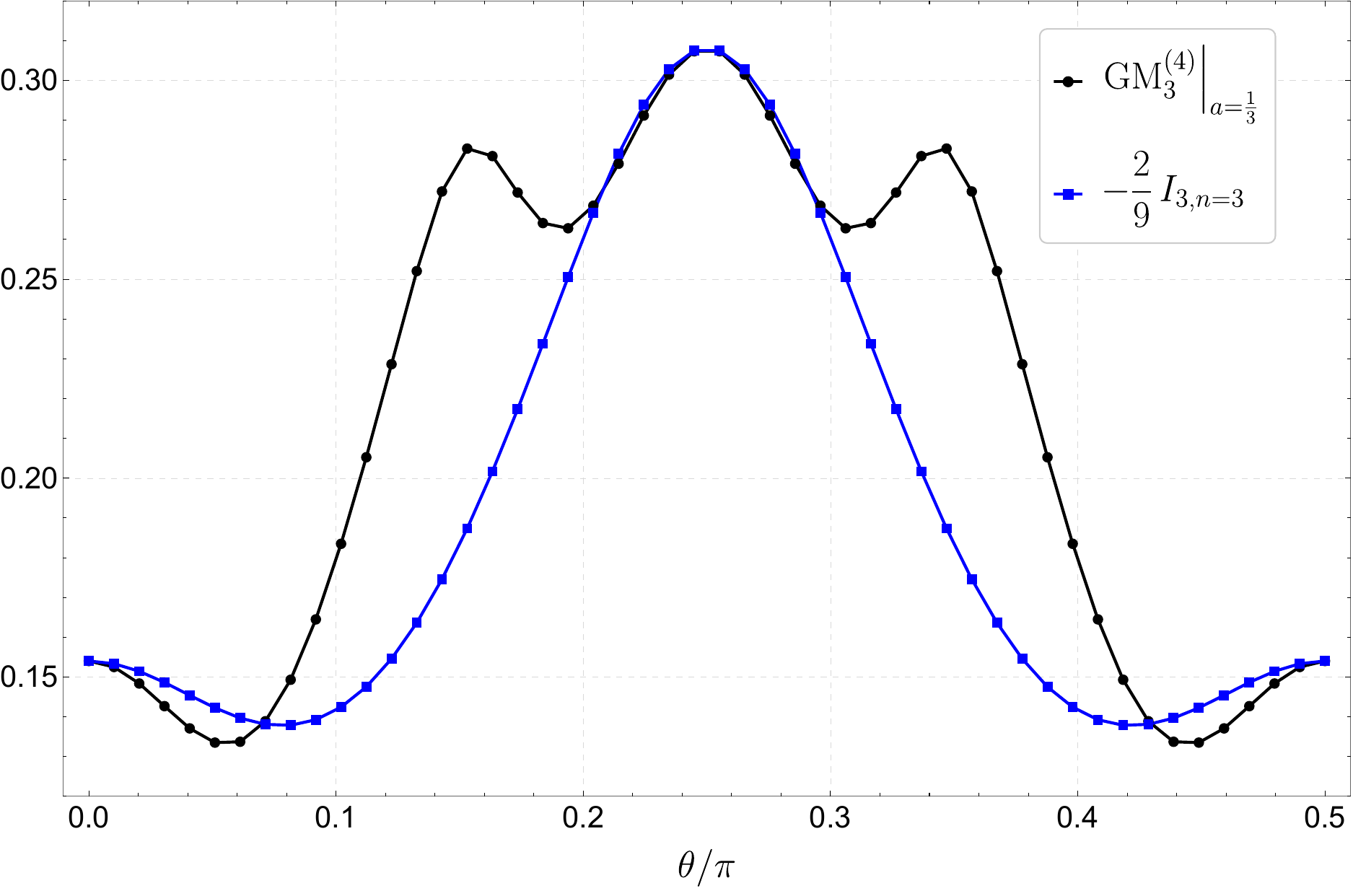}

\caption{
$\GM[\mathtt{q}=4]_n$ at $a=1/3$ and $- I_{3,n=2}/4$ for $|\Psi_{11}(\theta)\rangle$ at $n=2$ (top) and $-2 I_{3,n=3}/9$ at $n=3$ (bottom). The relations $\GM[\mathtt{q}=4]_{n=2}=-I_{3,n=2}/4$, \eqref{eq:GMn2}, and $\GM[\mathtt{q}=4]_{n=3}=-2I_{3,n=3}/9$, \eqref{eq:GMn3}, are satisfied at the stabilizer points $\theta=0,\pi/4,\pi/2$, but are generically violated.} 
\label{fig:Psi11}
\end{figure}

One can also consider an analogous state,
\begin{equation}
|\Psi_{11}(\theta)\rangle
=
\cos\theta\,|\psi_{00}\rangle
+
\sin\theta\,|\psi_{11}\rangle , \quad 0 \le \theta \le \frac{\pi}{2}.
\end{equation}
See Fig.~\ref{fig:Psi11} for the plots as a function of $\theta$. Although the constituent basis states differ between $|\Psi_{10}(\theta)\rangle$ and $|\Psi_{11}(\theta)\rangle$, the resulting values of $\GM[\mathtt{q}=4]_n$ and $I_{3,n}$ as functions of $\theta$ are identical to those shown in  Fig.~\ref{fig:Psi10} for both $n=2$ and $n=3$. Thus, as in the case of $|\Psi_{10}(\theta)\rangle$, for generic parameter $\theta$ in $\ket{\Psi_{11}(\theta)}$,  $\GM$ and $I_{3,n}$ are physically different quantities.

\subsubsection{Relative phase dependence for non-stabilizer family}

\begin{figure}[!t]
\centering

\includegraphics[width=0.95\columnwidth]{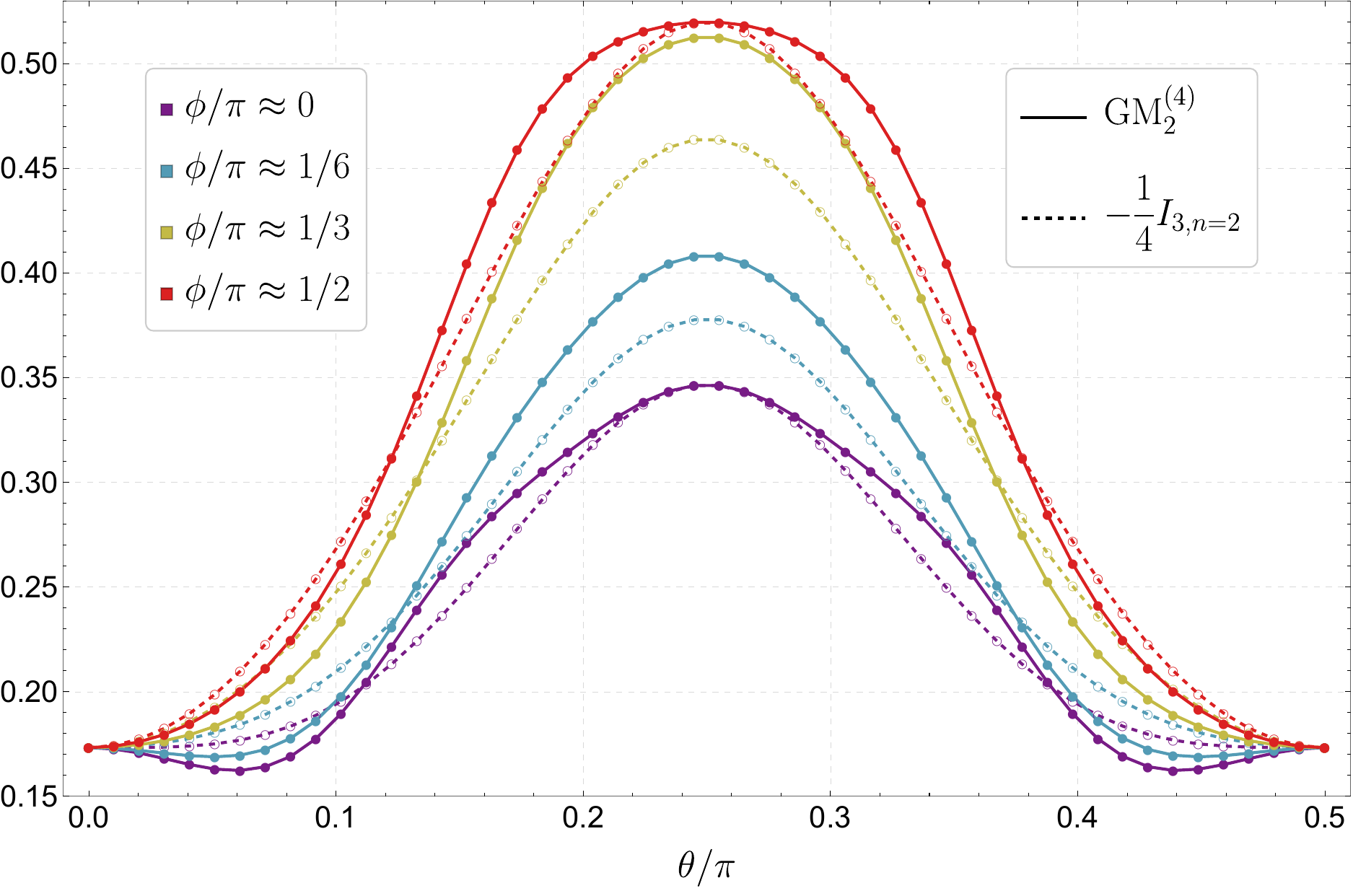}

\vspace{0.5cm}

\includegraphics[width=0.95\columnwidth]{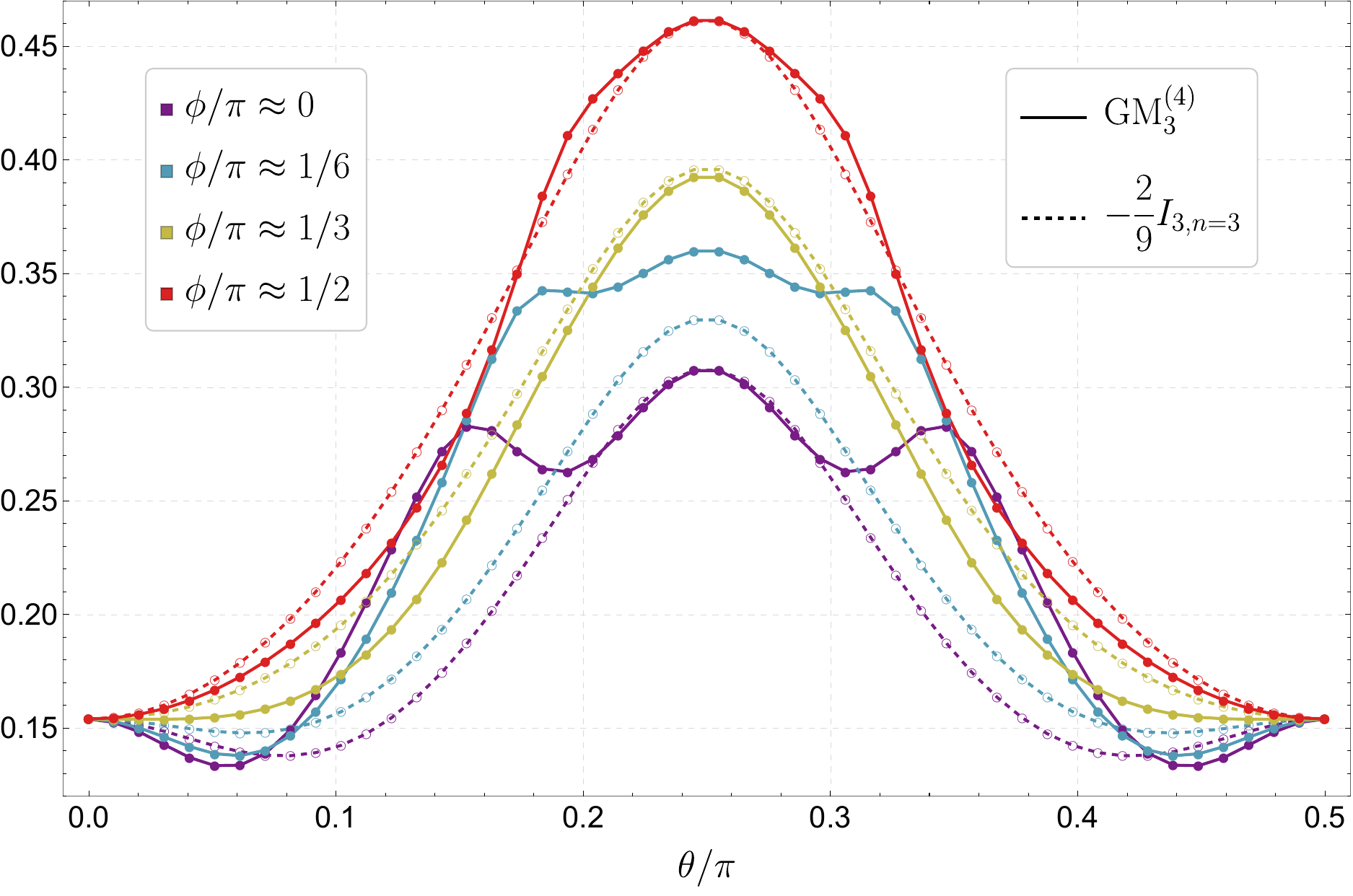}
\caption{
Plots of $\GM[\mathtt{q}=4]_n$ at $a=1/3$ together with $-I_{3,n=2}/4$ for $n=2$ (top) and $-2I_{3,n=3}/9$ for $n=3$ (bottom) for $|\Psi_{10}(\theta,\phi)\rangle$ as a function of $\theta$ for different values of the phase $\phi$. Only the range $0 \leq \phi \leq \pi/2$ is shown, since the curves are symmetric under the reflection $\phi \rightarrow \pi-\phi$. 
For $0<\phi<\pi/2$, the only remaining stabilizer points are $\theta=0$ and $\theta=\pi/2$.
}
\label{fig:Psi10_phase}
\end{figure}

We introduce a relative phase in \eqref{eq:Psi10_theta} as 
\begin{equation}
|\Psi_{10}(\theta,\phi)\rangle
=
\cos\theta\,|\psi_{00}\rangle
+
e^{i\phi}\sin\theta\,|\psi_{10}\rangle .
\end{equation}
See Fig.~\ref{fig:Psi10_phase} for the plots as a function of $\theta$ with different phases $\phi$. This plot shows that introducing a nonzero phase $0<\phi<\pi/2$ changes the set of stabilizer points: although $\theta=\pi/4$ is a stabilizer point for $\phi=0$ and $\phi=\pi/2$, for intermediate values $0<\phi<\pi/2$, it no longer satisfies the corresponding relation, \eqref{eq:GMn2} for $n=2$ and \eqref{eq:GMn3} for $n=3$. The remaining stabilizer points are $\theta=0$ and $\theta=\pi/2$. 
At all of these stabilizer points, the corresponding relation,
\eqref{eq:GMn2} for $n=2$ and \eqref{eq:GMn3} for $n=3$, is
satisfied. For generic non-stabilizer values of $(\theta,\phi)$,
however, GM and the corresponding multiple of $I_{3,n}$ do not agree.

\subsubsection{Restricted four-component family with real coefficients}

\begin{figure}[h]
\centering

\includegraphics[width=0.95\columnwidth]{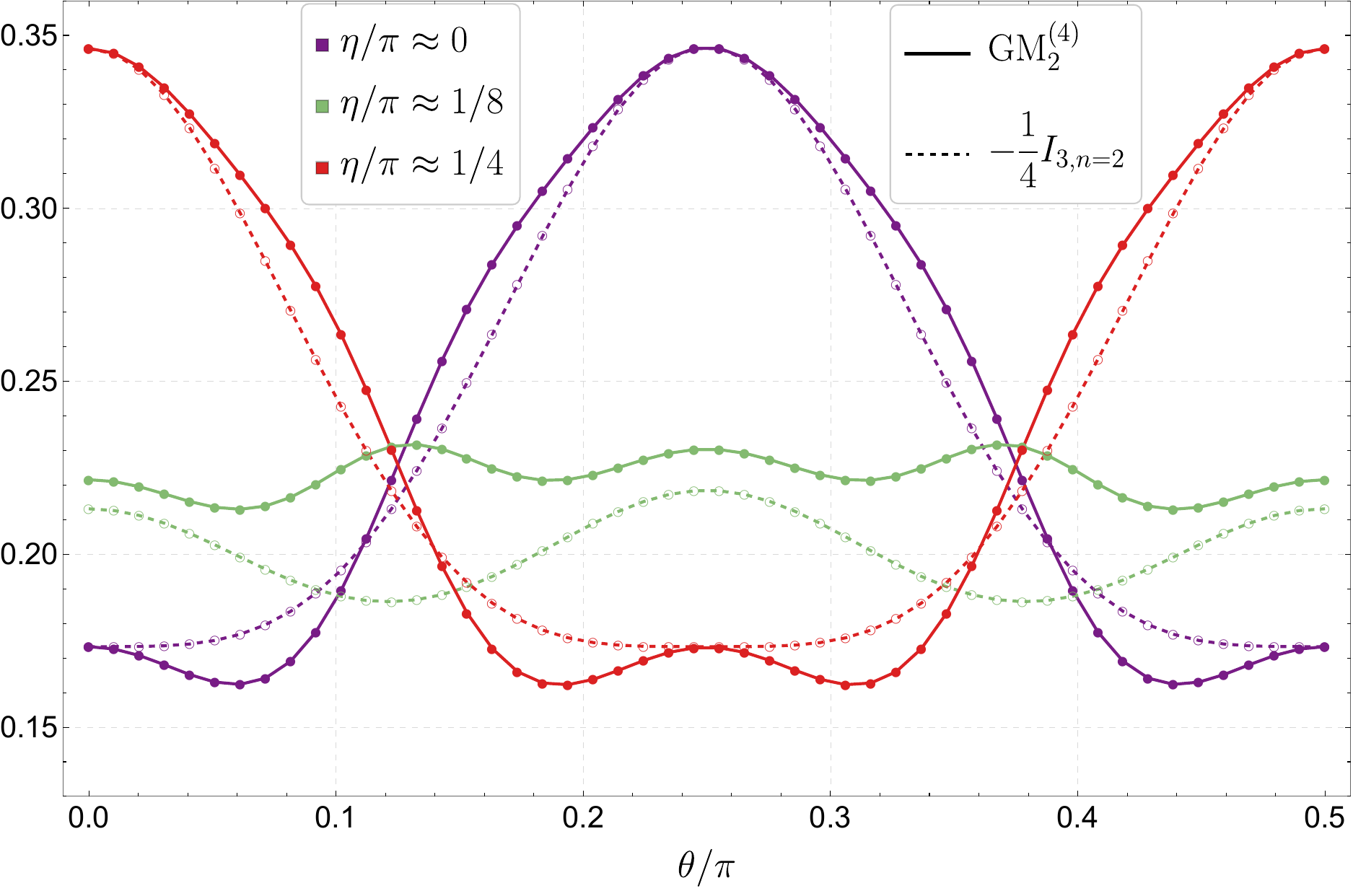}

\vspace{0.5cm}

\includegraphics[width=0.95\columnwidth]{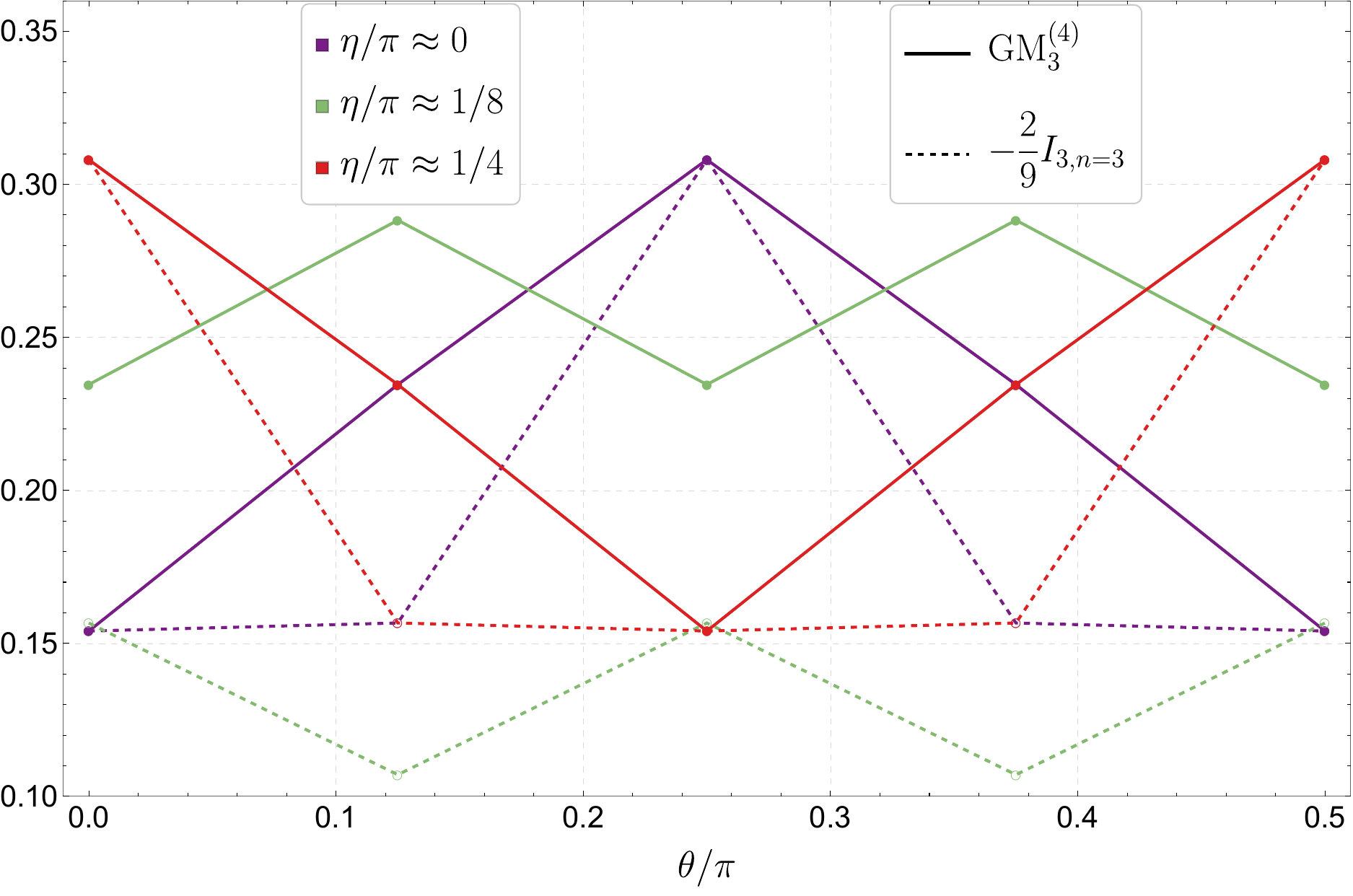}
\caption{
Plots of $\GM[\mathtt{q}=4]_n$ at $a=1/3$ together with $-I_{3,n=2}/4$ for $n=2$ (top) and $-2I_{3,n=3}/9$ for $n=3$ (bottom) for $|\Psi(\theta,\eta)\rangle$ as a function of $\theta$ for different values of $\eta$.
Owing to the higher computational cost for $n=3$, fewer values of $\theta$ are shown in the lower panel, while the same qualitative behavior as for $n=2$ is observed. 
Only the range $0 \leq \eta \leq \pi/4$ is shown, since the curves are symmetric under the reflection $\eta \rightarrow \pi/2-\eta$. 
The stabilizer points at $\theta=0,\pi/4,\pi/2$ are preserved only at the special values $\eta=0$ and $\eta=\pi/4$.
}
\label{fig:Psi_two_parameter}
\end{figure}

Next, we consider a restricted four-component family with real coefficients
\begin{equation}
\begin{aligned}
c_{00} &= \cos\theta\cos\eta,
\qquad
c_{10} = \sin\theta\cos\eta,
\\
c_{01} &= \cos\theta\sin\eta,
\qquad
c_{11} = \sin\theta\sin\eta.
\end{aligned}
\end{equation}
In other words,
\begin{align}
|\Psi(\theta,\eta)\rangle
&=
\cos\theta\cos\eta\,|\psi_{00}\rangle
+
\sin\theta\cos\eta\,|\psi_{10}\rangle
\nonumber \\
&\quad +
\cos\theta\sin\eta\,|\psi_{01}\rangle
+
\sin\theta\sin\eta\,|\psi_{11}\rangle .
\end{align}
This is a simple two-parameter subset of the general four-component ground-state manifold. See Fig.~\ref{fig:Psi_two_parameter} for the plot as a function of $\theta$ with different values of $\eta$. The figure indicates that, for intermediate values $0<\eta<\pi/4$, the stabilizer points at $\theta=0,\pi/4,\pi/2$ are lifted; consequently, the relations \eqref{eq:GMn2} for $n=2$ and \eqref{eq:GMn3} for $n=3$ are no longer satisfied at these points. 
These stabilizer points emerge only at the special values $\eta=0$ and $\eta=\pi/4$.

\section{GM for excited states}
\label{sec:local-excitations}

We next consider excited states.
First, let us act with a Pauli operator on a single edge, say $e_1$:
\begin{equation}
|\Psi^{(P)}\rangle
=
P_1|\Psi\rangle,
\qquad
P_1=X_1,Z_1,Y_1.
\end{equation}
These operations create excited states of the toric-code Hamiltonian.
However, they do not change the GM.

Indeed, $P_1$ acts only on the single edge $e_1$.
For the partitions considered in this work, $e_1$ belongs entirely to
one subsystem. Therefore, $P_1$ is a local unitary transformation with
respect to the relevant multipartition. It follows that
\begin{equation}
\GM[\mathtt{q}=4]_n\bigl[P_1|\Psi\rangle\bigr]
=
\GM[\mathtt{q}=4]_n\bigl[|\Psi\rangle\bigr],
\end{equation}
and similarly,
\begin{equation}
I_{3,n}\bigl[P_1|\Psi\rangle\bigr]
=
I_{3,n}\bigl[|\Psi\rangle\bigr].
\end{equation}
Consequently,
\begin{equation}
\left. \GM[\mathtt{q}=4]_n \right|_{a=1/3}-\lambda_n I_{3,n}
\end{equation}
is also invariant under a single-edge Pauli excitation.

More generally, the same conclusion holds for an arbitrary Pauli string
\begin{equation}
U_P
=
\prod_{i=1}^{8} P_i,
\qquad
P_i\in\{I,X_i,Y_i,Z_i\}.
\end{equation}
Although $U_P$ can create excitations on several edges and can have
support in more than one subsystem, it factorizes with respect to the
multipartition as
\begin{equation}
U_P
=
U_A\otimes U_B\otimes U_C\otimes U_D.
\end{equation}
It is therefore again a local unitary transformation. Hence,
\begin{align}
\GM[\mathtt{q}=4]_n\bigl[U_P|\Psi\rangle\bigr]
&=
\GM[\mathtt{q}=4]_n\bigl[|\Psi\rangle\bigr] \,, \\
I_{3,n}\bigl[U_P|\Psi\rangle\bigr]
&=
I_{3,n}\bigl[|\Psi\rangle\bigr].
\end{align}
Thus, neither single-edge Pauli excitations nor arbitrary Pauli-string
excitations can produce a nontrivial change in GM.

To obtain a nontrivial excited-state dependence, we next consider coherent superpositions of two locally excited states on different edges with a one-parameter weight built on the
one-parameter ground-state family \eqref{eq:Psi10_theta}; for each
value of $\theta$, we consider the normalized states
\begin{equation}
\begin{aligned}
	&|\Psi_{10}^{(P_1,P_2;\nu)}(\theta)\rangle\\
 & \qquad =
\frac{1}{\mathcal{N}_{P_1,P_2;\nu}(\theta)}
\left(\cos \nu \, P_1+ \sin \nu \, P_2\right)
|\Psi_{10}(\theta)\rangle,
\end{aligned}
\label{eq:coherent_excited_state}
\end{equation}
where $0 \leq \nu \leq \pi/2$ and 
\begin{equation}
\mathcal{N}_{P_1,P_2;\nu}(\theta)^2
=
\langle\Psi_{10}(\theta)|
\left(\cos \nu \, P_1+ \sin \nu \, P_2\right)^{2}
|\Psi_{10}(\theta)\rangle.
\label{eq:coherent_excited_normalization}
\end{equation}
Here, we focus on the three cases
\begin{equation}
(P_1,P_2)=(X_1,X_2), \, (X_1, Z_2), \, (Z_1,Z_2).
\label{eq:coherent_excitation_choices}
\end{equation}
Unlike a single Pauli excitation or an arbitrary Pauli string, these operators are non-unitary.
Moreover, their two terms act on different edges.  When these edges belong to different subsystems of the multipartition, the operator does not factorize into a product of operators acting separately on the individual subsystems.  The resulting state is therefore not, in general, related to the original ground state by a local unitary transformation.

For each choice in \eqref{eq:coherent_excitation_choices}, we compute $\GM[\mathtt{q}=4]_n$ at $a=1/3$ and $I_{3,n}$ for the $n=2$ case. This allows us to examine whether the relations found for stabilizer states remain valid after superposing Pauli-excited states. We note that the $n=3$ case is omitted here, as it exhibits qualitatively similar behavior to the $n=2$ case, similar to what was observed for the ground states in Sec.~\ref{subsec:general-ground-states}.

\begin{figure}[t]
\centering

\includegraphics[width=0.95\columnwidth]{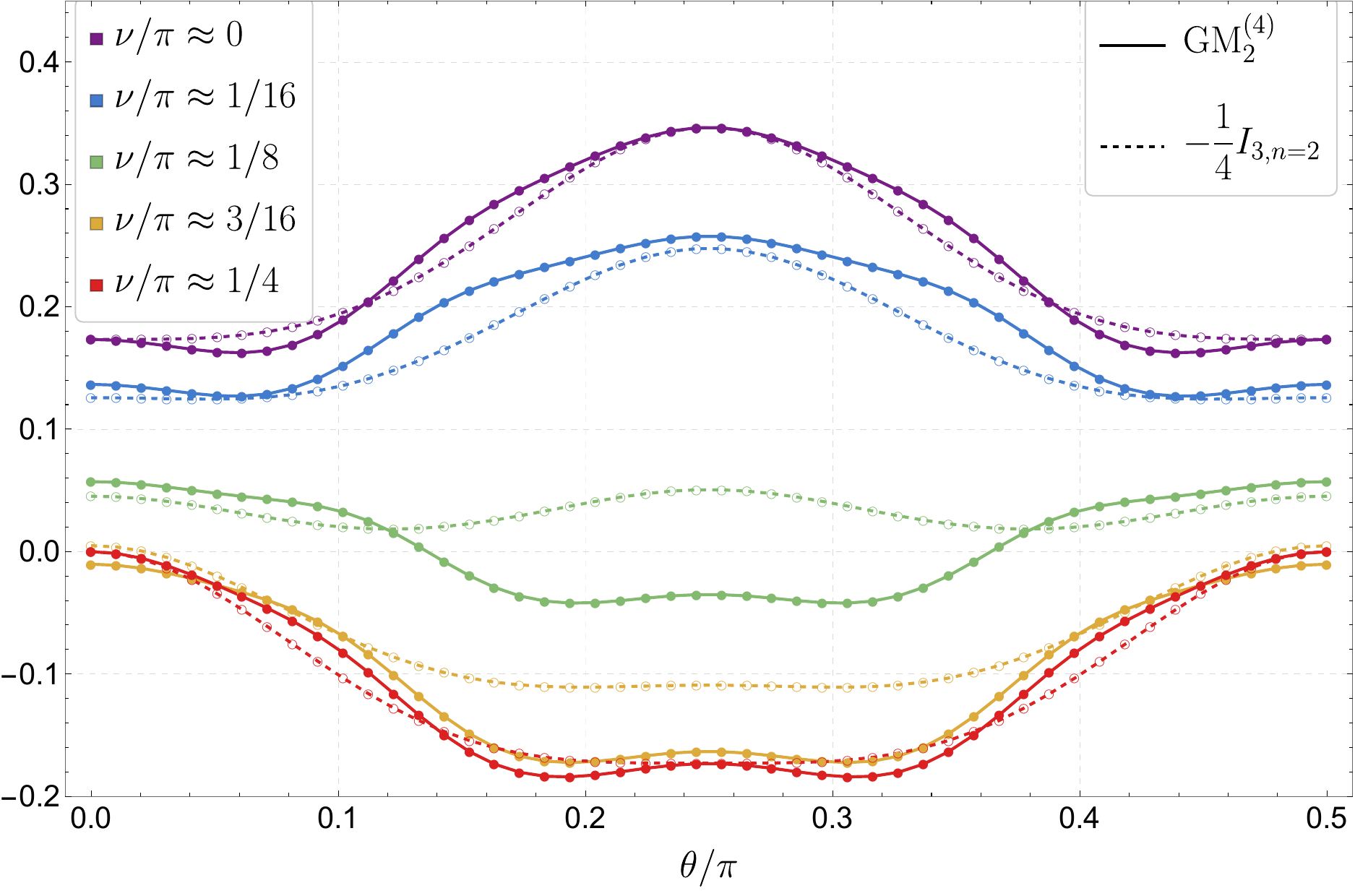}

\vspace{0.5cm}

\includegraphics[width=0.95\columnwidth]{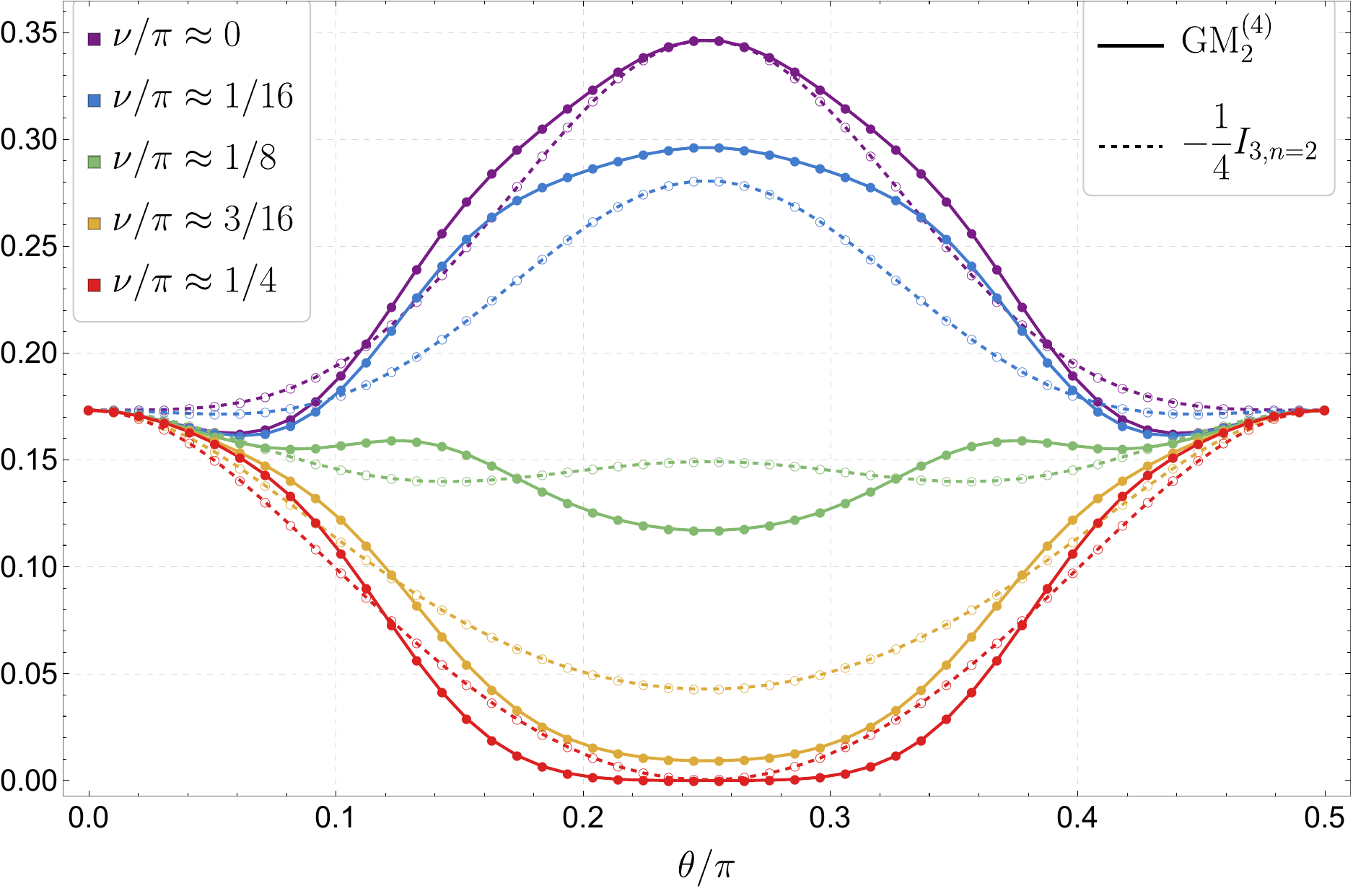}
\caption{
Plots of $\GM[\mathtt{q}=4]_2|_{a=1/3}$ and $-I_{3,n=2}/4$ for $|\Psi_{10}^{(P_1,P_2;\nu)}(\theta)\rangle$ with $(P_1,P_2)=(X_1,X_2)$ (top) and $(P_1,P_2)=(X_1, Z_2)$ (bottom) as a function of $\theta$ for different values of the mixing angle $\nu$.
 Only the range $0 \leq \nu \leq \pi/4$ is shown, since the curves are symmetric under the reflection $\nu \rightarrow \pi/2-\nu$. 
While for $(P_1,P_2)=(X_1,X_2)$ and $0<\nu<\pi/4$, there is no stabilizer point, for $(P_1,P_2)=(X_1,Z_2)$ and $0<\nu<\pi/4$, there are stabilizer points at $\theta=0,\pi/2$. 
}
\label{fig:Psi10_excited_X1X2-X1Z2}
\end{figure}

\begin{figure}[t]
\centering

\includegraphics[width=0.95\columnwidth]{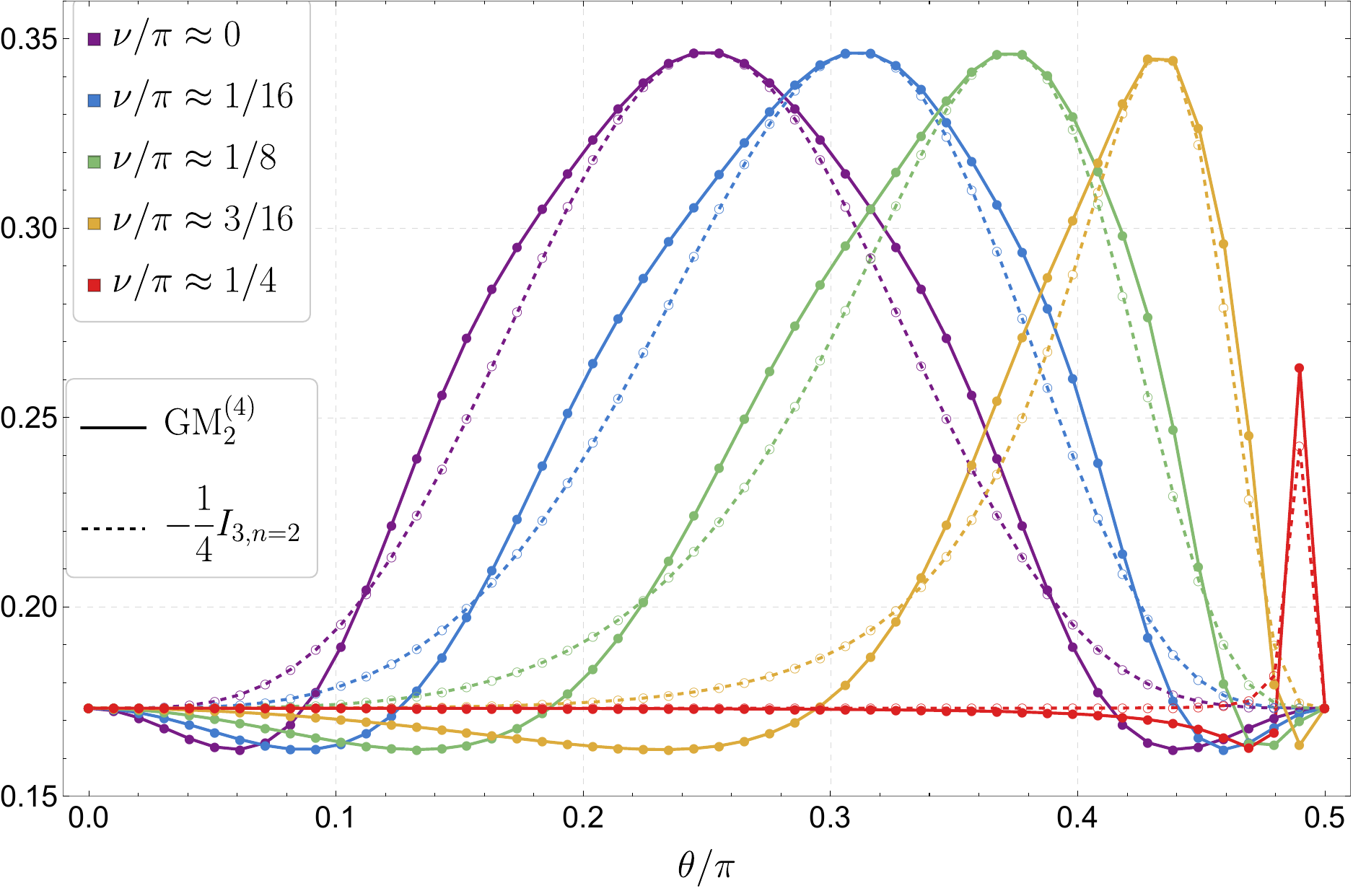}
\caption{
Plots of $\GM[\mathtt{q}=4]_2|_{a=1/3}$ and $-I_{3,n=2}/4$ for $|\Psi_{10}^{(P_1,P_2;\nu)}(\theta)\rangle$ with $(P_1,P_2)=(Z_1,Z_2)$ as a function of $\theta$ for different values of the mixing angle $\nu$.
Only the range $0 \leq \nu \leq \pi/4$ is shown, since the curves are symmetric under the reflection $\nu \rightarrow \pi/2-\nu$. The stabilizer points at $\theta=0$ and $\pi/2$ remain fixed for all $\nu$, while the intermediate stabilizer point shifts to $\theta=\pi/4+\nu$ as derived from (\ref{eq:deformed-angle}). At $\nu \approx \pi/4$ (more precisely, $\nu = 0.244\dots \pi$), the state is close to a stabilizer state over almost all of the displayed range of $\theta$.
}
\label{fig:Psi10_excited_Z1Z2}
\end{figure}

The results are shown in Fig.~\ref{fig:Psi10_excited_X1X2-X1Z2} and Fig.~\ref{fig:Psi10_excited_Z1Z2}.
 As seen in the figures, there is a clear qualitative difference between the cases. The $(X_1, X_2)$ and $(X_1, Z_2)$ cases generate nontrivial entanglement features that depend strongly on the mixing angle $\nu$.
 In contrast, the plot for the $(Z_1, Z_2)$ case largely retains the profile of the unexcited ground state \eqref{eq:Psi10_theta}, merely appearing deformed or shifted relative to the top panel of Fig.~\ref{fig:Psi11}.

This difference can be understood by analyzing the algebraic structure of the coherent excitation operators. First, for the $(Z_1, Z_2)$ case, the excitation operator can be written as
\begin{equation}
\begin{aligned}
    \cos \nu \, Z_1 + \sin \nu \, Z_2= Z_1 (\cos \nu + \sin \nu \, W_{x}^{Z}),
\label{eq:Z1Z2_factorization}
\end{aligned}
\end{equation}
where $W_{x}^{Z}=Z_1 Z_2$, \eqref{eq:loop_operators}.
Because $Z_1$ acts as a local unitary, it does not change $\GM[\mathtt{q}=4]_n$ or $I_{3,n}$ due to the local unitary invariance of these quantities.
Thus, the nontrivial change in the entanglement structure is entirely governed by the operator $\cos \nu + \sin \nu \, W_{x}^{Z}$. Crucially, the basis states $|\psi_{00}\rangle$ and $|\psi_{10}\rangle$ composing the one-parameter ground-state family $|\Psi_{10}(\theta)\rangle$, \eqref{eq:Psi10_theta}, are eigenstates of the loop operator $W_{x}^{Z}$. One can explicitly verify this fact by using the explicit forms of $|\psi_{00}\rangle$ and $|\psi_{10}\rangle$ given by \eqref{eq:psi00} and \eqref{eq:psi10} respectively. More generally, any operator that acts diagonally on the basis ground states with distinct eigenvalues will merely reweight their one-parameter superposition in this manner. In the present topological model, $W_{x}^{Z}$ naturally plays this role because the non-contractible loop operator $W^Z_x$ distinguishes these ground-state sectors.
Consequently, applying this operator does not generate a new orthogonal state; it merely reweights the original superposition coefficients. Thus, $\GM[\mathtt{q}=4]_n$ and $I_{3,n}$ for the resulting excited state $|\Psi_{10}^{(Z_1,Z_2;\nu)}(\theta)\rangle$ are the same as those of an unexcited ground state $|\Psi_{10}(\tilde{\theta})\rangle$ belonging to the same family, but with a modified effective parameter $\tilde{\theta}(\theta,\nu)$ given by
\begin{equation}
	\begin{aligned}
		\cos \tilde{\theta} &= \frac{\cos\theta (\cos\nu + \sin\nu)}{\sqrt{1 + \sin(2\nu)\cos(2\theta)}},\\
	 \sin \tilde{\theta} &= \frac{\sin\theta (\cos\nu - \sin\nu)}{\sqrt{1 + \sin(2\nu)\cos(2\theta)}}.
	\end{aligned}\label{eq:deformed-angle}
\end{equation}
Here, we note that the denominators in \eqref{eq:deformed-angle} vanish at $\nu=\pi/4$ and $\theta=\pi/2$. In Fig.~\ref{fig:Psi10_excited_Z1Z2}, we avoid this singularity by considering a slightly smaller value, $\nu/\pi = 0.244\cdots \approx 1/4$.

According to \eqref{eq:deformed-angle}, while the stabilizer points at $\theta=0$ and $\pi/2$ remain fixed independently of $\nu$, the intermediate stabilizer point originally located at $\theta=\pi/4$ is shifted to $\theta=\pi/4+\nu$. Furthermore, for strictly $\nu=\pi/4$, the state completely collapses to the $\tilde{\theta}=0$ stabilizer state for all $\theta$. The slight deviation near $\theta=\pi/2$ visible in Fig.~\ref{fig:Psi10_excited_Z1Z2} simply reflects the fact that the plotted curve corresponds to $\nu \approx \pi/4$ rather than this exact limit.
This explains why Fig.~\ref{fig:Psi10_excited_Z1Z2} shows no fundamentally new entanglement structures outside the original one-parameter family state, \eqref{eq:Psi10_theta}, but instead displays essentially the same profile as the top panel of Fig.~\ref{fig:Psi11} with some shift or deformation. 

On the other hand, for the $(X_1, X_2)$ and $(X_1, Z_2)$ cases, factoring out the local unitary $X_1$ gives the operators $\cos \nu + \sin \nu\, X_1X_2$ and $\cos \nu + \sin \nu\, X_1Z_2$, respectively. Unlike $Z_1Z_2$, these bilinear Pauli operators do not preserve the ground-state subspace spanned by $\{|\psi_{00}\rangle,|\psi_{10}\rangle\}$. Instead, they generate components outside this subspace. Moreover, their expectation values in $|\Psi_{10}(\theta)\rangle$ vanish, so these components are orthogonal to the original state. Therefore, the resulting states cannot be absorbed into a simple reparametrization of $\theta$.
This is why the $(X_1,X_2)$ and $(X_1,Z_2)$ excitations produce genuinely new, $\nu$-dependent entanglement structures, as observed in Fig.~\ref{fig:Psi10_excited_X1X2-X1Z2}.

\section{Discussions}
\label{discussions}

In this paper, we studied the genuine multi-entropy (GM) in the toric code by considering four-partite decompositions of the torus. We conjectured that for any stabilizer state, the $\mathtt{q} \,(\ge 4)$-partite GM reduces to a combination of entropies involving at most $\mathtt{q}-2$ parties, unless the R\'enyi index $n \geq \mathtt{q}$. We explicitly checked this for $\mathtt{q} = 4$ where both the $n = 2$ and $n = 3$ GM reduce to the tripartite information $I_3$. We then turned to Kitaev's toric code and provided evidence that the $n = 4$ GM is a topological invariant of the underlying theory within the precisely defined deformation class of Sec.~\ref{KitaevPreskillpartitions}.
As the $n = 4$ GM does not collapse to the tripartite information, it probes TQFT data not visible to $I_3$. 
However, for non-stabilizer states, the collapse of the genuine multi-entropy to the tripartite information generically fails, demonstrating that $\GM[\mathtt{q=4}]_n$ detects genuinely four-partite correlations beyond those captured by $I_{3,n}$ and the topological entanglement entropy.

This raises several interesting questions. First, why does the $n < \mathtt{q}$ GM reduce to a linear combination of smaller-partite entropies? When $n = 2$, there's a simple explanation for this owing to the Coxeter structure of the permutations that define the $n = 2$ R\'enyi multi-entropy. As argued in \cite{Akella:2026xza}, there is an explicit counting argument using the Coxeter structure that reproduces the collapse of the $n = 2$ GM conjectured in \cite{Iizuka:2025pqq}. Whether there is a similar counting argument for $n = 3$ case is an open question.

In the toric code, the tripartite information has a clean TQFT interpretation for the Kitaev--Preskill partition. It calculates the total quantum dimension $\mathcal{D}$ of the underlying TQFT. While we numerically argued that the $n = 4$ GM is a topological invariant that does not collapse to $I_3$, we haven't explained what it is in terms of the underlying TQFT data. Recently, \cite{DelZotto:2026fpw} showed that any genuine multipartite signal, including the GM, evaluated for a Kitaev--Preskill type partition for the ground state of a Levin-Wen model, is the TQFT partition function on the graph-encoded manifold. This is essentially a consequence of the replica trick. The novelty of the signal construction, however, is that it eliminates all the UV divergences and gives a finite TQFT answer. From this argument, it must be that the $n = 4$ GM is simply computing the toric-code TQFT partition function on the gem corresponding to the $n = 4$ R\'enyi multi-entropy. Figuring out the topology of the $n = 4$ gem and working this calculation out is left to future work.

A related and perhaps more interesting question is to calculate the $n = 4$ GM for the double semion theory \cite{Levin:2006zz} and compare it to the toric code. The toric code TQFT and the double semion TQFT have the same total quantum dimension $\mathcal{D}$, but are inequivalent TQFTs. If the $n = 4$ GM probes the deeper TQFT structure, then the obvious place to look is the double semion theory. The Levin-Wen model of the double semion theory, however, is not a simple Pauli stabilizer error-correcting code like the toric code. This is because of factors of $i$ in the braiding rules of the double semion theory that ruin the stabilizer structure. We therefore cannot apply the numerical techniques from \cite{Akella:2026xza} to calculate the GM efficiently. However, \cite{Ellison:2021vth} realize the double semion in the stabilizer framework starting with a $\mathbb{Z}_4$ toric code and making certain two-body measurements to condense an emergent boson. 
A natural next step is to take their setup and work out the $n = 4$ GM of the double semion theory.
These observations suggest a promising direction in which genuine multi-entropy can be used to distinguish and characterize topological phases beyond the information contained in the total quantum dimension. We hope to report further progress on these questions in the near future.

\acknowledgments
S.A. is supported by the Department of Atomic Energy, Government of India, under Project Identification No. RTI 4002, and the Infosys Endowment for the study of the Quantum Structure of Spacetime. The work of N.I. was supported in part by MEXT KAKENHI Grant-in-Aid for Transformative Research Areas A “Extreme Universe” No. 21H05184. The work of N.I. was also supported in part by NSTC of Taiwan Grant Number 114-2112-M-007-025-MY3.  The work of A.M. was supported by JSPS KAKENHI Grant Number JP26KJ0186. We used Claude (Anthropic) to assist with numerical calculations and independent checks of some results. All AI-assisted numerical results were independently checked and verified by the authors.

\bibliography{refs}

\appendix 

\section{Replica definition of multi-entropy}
\label{app:Mdefinition}

In this appendix, we give the replica definition of the R\'enyi
multi-entropy used in the main text.  Consider a pure state
$\ket{\psi}$ on a $\mathtt{q}$-partite Hilbert space
\begin{equation}
\mathcal{H}
=
\mathcal{H}_{A_1}\otimes\mathcal{H}_{A_2}
\otimes\cdots\otimes\mathcal{H}_{A_{\mathtt{q}}}.
\end{equation}
The R\'enyi-$n$ $\mathtt{q}$-partite multi-entropy is defined by
\begin{align}
S_n^{(\mathtt{q})}(A_1:\cdots:A_{\mathtt{q}})
=
\frac{1}{1-n}\frac{1}{n^{\mathtt{q}-2}}
\log\left[
\frac{Z_n^{(\mathtt{q})}}{\left(Z_1^{(\mathtt{q})}\right)^{n^{\mathtt{q}-1}}}
\right],
\label{eq:multi_entropy_replica_definition}
\end{align}
where
\begin{align}
Z_n^{(\mathtt{q})}
=
\bra{\psi}^{\otimes n^{\mathtt{q}-1}}
\Sigma_1(g_1)\Sigma_2(g_2)\cdots\Sigma_{\mathtt{q}}(g_{\mathtt{q}})
\ket{\psi}^{\otimes n^{\mathtt{q}-1}}.
\label{eq:multi_entropy_partition_function}
\end{align}
Here $\Sigma_i(g_i)$ acts on the replicas of subsystem $A_i$ according
to the permutation $g_i$.

The replicas are labeled by lattice points
\begin{equation}
(x_1,\ldots,x_{\mathtt{q}-1}),
\qquad
x_i=1,\ldots,n,
\end{equation}
on a $(\mathtt{q}-1)$-dimensional periodic hypercubic lattice.  The permutations
are defined by
\begin{align}
g_i\cdot(x_1,\ldots,x_i,\ldots,x_{\mathtt{q}-1})
=
(x_1,\ldots,x_i+1,\ldots,x_{\mathtt{q}-1}),
\label{eq:multi_entropy_permutation}
\end{align}
for $i=1,\ldots,\mathtt{q}-1$, with $x_i=n+1$ identified with $x_i=1$, while
\begin{equation}
g_{\mathtt{q}}=e.
\end{equation}
For $\mathtt{q}=2$, this construction reduces to the ordinary R\'enyi entropy,
\begin{equation}
S_n^{(2)}(A:B)=S_n(A)=S_n(B).
\end{equation}

\section{An explicit ground-state basis for the $2 \times 2$ lattice toric code}
\label{Apptoric2by2}

Let us work in the \(Z\)-basis and denote a basis vector by
\begin{equation}
|s_1s_2s_3s_4s_5s_6s_7s_8\rangle  =
\left|
\underbrace{s_1s_2s_3s_4}_{
\begin{subarray}{c}
e_1,e_2,e_3,e_4\\
\text{horizontal}
\end{subarray}}
\underbrace{s_5s_6s_7s_8}_{
\begin{subarray}{c}
e_5,e_6,e_7,e_8\\
\text{vertical}
\end{subarray}}
\right\rangle .
\end{equation}
where \(s_i=0,1\) is the computational-basis value on the edge qubit \(e_i\).
In our convention for the \(L=2\) lattice, the first \(L^2=4\)
qubits,
$e_1,e_2,e_3,e_4$
are horizontal edges, while the remaining \(L^2=4\) qubits,
$e_5,e_6,e_7,e_8$
are vertical edges.

The star and plaquette operators are explicitly
\begin{equation}
\begin{aligned}
A_{11} &= X_1X_2X_5X_7 \,, \quad  
A_{12} = X_1X_2X_6X_8,\\
A_{21} &= X_3X_4X_5X_7 \,, \quad 
A_{22} = X_3X_4X_6X_8\,,\\
B_{11} &= Z_1Z_3Z_5Z_6 \,, \quad 
\,\, \, B_{12} = Z_2Z_4Z_5Z_6,\\
B_{21} &= Z_1Z_3Z_7Z_8 \,, \quad
\,\, \,B_{22} = Z_2Z_4Z_7Z_8 \,.
\end{aligned}
\end{equation}

The four-dimensional ground-state subspace is given by
\begin{align}
& |\psi_{00}\rangle
=\frac{1}{\sqrt{8}}\big(
|00000000\rangle \label{eq:psi00}
+|00001111\rangle  \nonumber \\
& \quad +|00110101\rangle
+|00111010\rangle
+|11000101\rangle  \\
& \quad +|11001010\rangle
+|11110000\rangle
+|11111111\rangle
\big), \nonumber  \\
&|\psi_{10}\rangle
=\frac{1}{\sqrt{8}}\big(
|01010000\rangle \label{eq:psi10}
+|01011111\rangle \nonumber \\
& \quad +|01100101\rangle
+|01101010\rangle 
+|10010101\rangle \\
& \quad +|10011010\rangle
+|10100000\rangle
+|10101111\rangle
\big), \nonumber \\
& |\psi_{01}\rangle
=\frac{1}{\sqrt{8}}\big(
|00000011\rangle
+|00001100\rangle \nonumber \\
& \quad +|00110110\rangle
+|00111001\rangle 
+|11000110\rangle \\
& \quad +|11001001\rangle
+|11110011\rangle
+|11111100\rangle
\big), \nonumber \\
& |\psi_{11}\rangle
=\frac{1}{\sqrt{8}}\big(
|01010011\rangle
+|01011100\rangle \nonumber \\
&\quad +|01100110\rangle
+|01101001\rangle
+|10010110\rangle \\
& \quad +|10011001\rangle
+|10100011\rangle
+|10101100\rangle
\big). \nonumber
\end{align}

These four states satisfy
\begin{equation}
A_v|\psi_{\alpha\beta}\rangle=|\psi_{\alpha\beta}\rangle,
\qquad
B_p|\psi_{\alpha\beta}\rangle=|\psi_{\alpha\beta}\rangle
\end{equation}
for all vertices $v$ and plaquettes $p$. They span the four-dimensional
ground-state subspace of the $L=2$ toric code on the torus. The labels $\alpha,\beta=0,1$ distinguish the four topological sectors,
or equivalently the eigenvalues of two independent non-contractible loop
operators as given by \eqref{eq:defofalphabeta}.

%

\section{From the checkerboard partition to the Kitaev--Preskill disk partition}
\label{app:KPdisk}

In this appendix we show explicitly that the four-partite ``checkerboard''
edge partition used in the main text \eqref{checkerboard} is connected, by a sequence of local
single-edge deformations that never cross a non-contractible cycle, to a
Kitaev--Preskill disk partition, and that $I_3$ stays equal to
$-\log 2$ along the entire sequence. This makes concrete the statement that
$I_3=-\gamma=-\log\mathcal{D}$ for the partition used in the main text.

\subsection{Setup}
\label{app:KPdisk-setup}

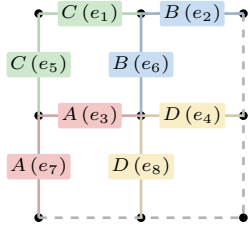
\begin{figure}[t]
\centering
\begin{tikzpicture}[scale=1.35,line width=1.0pt]
  \foreach \x in {0,1,2}\foreach \y in {0,1,2}{\fill (\x,\y) circle (1.2pt);}
  \draw[gray!60,dashed] (0,0)--(1,0)--(2,0);
  \draw[gray!60,dashed] (2,2)--(2,1)--(2,0);
  \draw[gray!60,dashed] (2,2)--(1,2);
  \draw[Ccol!90!black] (0,2)--(1,2); \node[fill=Ccol,rounded corners=1pt,inner sep=1.3pt] at (0.5,2){\scriptsize $C\,(e_1)$};
  \draw[Bcol!80!black] (1,2)--(2,2); \node[fill=Bcol,rounded corners=1pt,inner sep=1.3pt] at (1.5,2){\scriptsize $B\,(e_2)$};
  \draw[Acol!85!black] (0,1)--(1,1); \node[fill=Acol,rounded corners=1pt,inner sep=1.3pt] at (0.5,1){\scriptsize $A\,(e_3)$};
  \draw[Dcol!85!black] (1,1)--(2,1); \node[fill=Dcol,rounded corners=1pt,inner sep=1.3pt] at (1.5,1){\scriptsize $D\,(e_4)$};
  \draw[Ccol!90!black] (0,2)--(0,1); \node[fill=Ccol,rounded corners=1pt,inner sep=1.3pt] at (0,1.5){\scriptsize $C\,(e_5)$};
  \draw[Bcol!80!black] (1,2)--(1,1); \node[fill=Bcol,rounded corners=1pt,inner sep=1.3pt] at (1,1.5){\scriptsize $B\,(e_6)$};
  \draw[Acol!85!black] (0,1)--(0,0); \node[fill=Acol,rounded corners=1pt,inner sep=1.3pt] at (0,0.5){\scriptsize $A\,(e_7)$};
  \draw[Dcol!85!black] (1,1)--(1,0); \node[fill=Dcol,rounded corners=1pt,inner sep=1.3pt] at (1,0.5){\scriptsize $D\,(e_8)$};
\end{tikzpicture}
\caption{The $2\times2$ toric lattice with the partition \eqref{eq:Cpart0}.
Solid coloured edges are the eight independent qubits; dashed grey edges are
their periodic copies. Each party collects the horizontal and vertical edge
of one vertex: $C=\{e_1,e_5\}$, $B=\{e_2,e_6\}$, $A=\{e_3,e_7\}$,
$D=\{e_4,e_8\}$. Relative to the main text, $A$ and $C$ are interchanged.}
\label{fig:Appfig1}
\end{figure}

We work on the $2\times2$ toric lattice of Fig.~\ref{Fig:2by2lattice}, with
the edge labelling $e_1,\dots,e_8$ introduced there. We start from the four-partite
partition shown in Fig.~\ref{fig:Appfig1}, which is  checkerboard edge partition:
\begin{equation}
\label{eq:Cpart0}
\begin{aligned}
C&=\{e_1,e_5\},\qquad B=\{e_2,e_6\},\\
A&=\{e_3,e_7\},\qquad D=\{e_4,e_8\}.
\end{aligned}
\end{equation}
Each party is the pair of edges (one horizontal, one vertical) emanating
from one vertex, so $A\sqcup B\sqcup C\sqcup D=\{e_1,\dots,e_8\}$. Relative
to the main-text assignment, \eqref{checkerboard}, $A=\{e_1,e_5\}$, $C=\{e_3,e_7\}$, the labels $A$
and $C$ are interchanged here. Since $I_3(A{:}B{:}C)$ is symmetric in its
three arguments, this does not affect any value below.

The reference stabilizer ground state is $\ket{\psi_{\alpha \beta}}$ (see also App.~\ref{Apptoric2by2})

\subsection{Step-by-step deformation}
\label{app:KPdisk-deform}

\begin{figure}[t]
\centering
\begin{tikzpicture}[scale=1.35,line width=1.0pt]
  \foreach \x in {0,1,2}\foreach \y in {0,1,2}{\fill (\x,\y) circle (1.2pt);}
  \draw[gray!60,dashed] (0,0)--(1,0)--(2,0);
  \draw[gray!60,dashed] (2,2)--(2,1)--(2,0);
  \draw[gray!60,dashed] (2,2)--(1,2);
  \draw[Ccol!90!black] (0,2)--(1,2); \node[fill=Ccol,rounded corners=1pt,inner sep=1.3pt] at (0.5,2){\scriptsize $C\,(e_1)$};
  \draw[Bcol!80!black] (1,2)--(2,2); \node[fill=Bcol,rounded corners=1pt,inner sep=1.3pt] at (1.5,2){\scriptsize $B\,(e_2)$};
  \draw[Acol!85!black] (0,1)--(1,1); \node[fill=Acol,rounded corners=1pt,inner sep=1.3pt] at (0.5,1){\scriptsize $A\,(e_3)$};
  \draw[Dcol!85!black] (1,1)--(2,1); \node[fill=Dcol,rounded corners=1pt,inner sep=1.3pt] at (1.5,1){\scriptsize $D\,(e_4)$};
  \draw[Ccol!90!black] (0,2)--(0,1); \node[fill=Ccol,rounded corners=1pt,inner sep=1.3pt] at (0,1.5){\scriptsize $C\,(e_5)$};
  \draw[Bcol!80!black] (1,2)--(1,1); \node[fill=Bcol,rounded corners=1pt,inner sep=1.3pt] at (1,1.5){\scriptsize $B\,(e_6)$};
  \draw[Dcol!85!black] (0,1)--(0,0); \node[fill=Dcol,rounded corners=1pt,inner sep=1.3pt] at (0,0.5){\scriptsize $D\,(e_7)$};
  \draw[Dcol!85!black] (1,1)--(1,0); \node[fill=Dcol,rounded corners=1pt,inner sep=1.3pt] at (1,0.5){\scriptsize $D\,(e_8)$};
\end{tikzpicture}
\caption{The partition \eqref{eq:Cpart1}:
$C=\{e_1,e_5\}$, $B=\{e_2,e_6\}$, $A=\{e_3\}$, $D=\{e_4,e_7,e_8\}$.
Compared with Fig.~\ref{fig:Appfig1}, the edge $e_7$ is assigned to $D$
rather than $A$.}
\label{fig:Appfig2}
\end{figure}
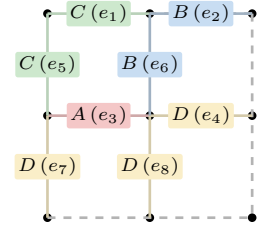

The first step deforms Fig.~\ref{fig:Appfig1} by reassigning edge $e_7$ from
$A$ to $D$,
\begin{equation}
\label{eq:Cpart1}
\begin{aligned}
C&=\{e_1,e_5\},\qquad B=\{e_2,e_6\},\\
A&=\{e_3\},\qquad D=\{e_4,e_7,e_8\}.
\end{aligned}
\end{equation}
The corresponding partition is Fig.~\ref{fig:Appfig2}. 

\begin{figure}[t]
\centering
\begin{tikzpicture}[scale=1.35,line width=1.0pt]
  \foreach \x in {0,1,2}\foreach \y in {0,1,2}{\fill (\x,\y) circle (1.2pt);}
  \draw[gray!60,dashed] (0,0)--(1,0)--(2,0);
  \draw[gray!60,dashed] (2,2)--(2,1)--(2,0);
  \draw[gray!60,dashed] (2,2)--(1,2);
  \draw[Ccol!90!black] (0,2)--(1,2); \node[fill=Ccol,rounded corners=1pt,inner sep=1.3pt] at (0.5,2){\scriptsize $C\,(e_1)$};
  \draw[Dcol!85!black] (1,2)--(2,2); \node[fill=Dcol,rounded corners=1pt,inner sep=1.3pt] at (1.5,2){\scriptsize $D\,(e_2)$};
  \draw[Acol!85!black] (0,1)--(1,1); \node[fill=Acol,rounded corners=1pt,inner sep=1.3pt] at (0.5,1){\scriptsize $A\,(e_3)$};
  \draw[Dcol!85!black] (1,1)--(2,1); \node[fill=Dcol,rounded corners=1pt,inner sep=1.3pt] at (1.5,1){\scriptsize $D\,(e_4)$};
  \draw[Ccol!90!black] (0,2)--(0,1); \node[fill=Ccol,rounded corners=1pt,inner sep=1.3pt] at (0,1.5){\scriptsize $C\,(e_5)$};
  \draw[Bcol!80!black] (1,2)--(1,1); \node[fill=Bcol,rounded corners=1pt,inner sep=1.3pt] at (1,1.5){\scriptsize $B\,(e_6)$};
  \draw[Dcol!85!black] (0,1)--(0,0); \node[fill=Dcol,rounded corners=1pt,inner sep=1.3pt] at (0,0.5){\scriptsize $D\,(e_7)$};
  \draw[Dcol!85!black] (1,1)--(1,0); \node[fill=Dcol,rounded corners=1pt,inner sep=1.3pt] at (1,0.5){\scriptsize $D\,(e_8)$};
\end{tikzpicture}
\caption{The partition \eqref{eq:Cpart2}:
$C=\{e_1,e_5\}$, $B=\{e_6\}$, $A=\{e_3\}$, $D=\{e_2,e_4,e_7,e_8\}$.
Compared with Fig.~\ref{fig:Appfig2}, the edge $e_2$ is assigned to $D$
rather than $B$.}
\label{fig:Appfig3}
\end{figure}
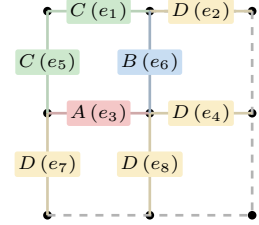

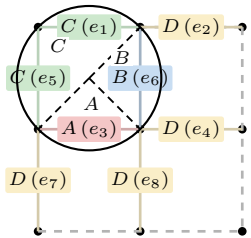
\begin{figure}[t]
\centering
\begin{tikzpicture}[scale=1.35,line width=1.0pt]
  \foreach \x in {0,1,2}\foreach \y in {0,1,2}{\fill (\x,\y) circle (1.2pt);}
  \draw[gray!60,dashed] (0,0)--(1,0)--(2,0);
  \draw[gray!60,dashed] (2,2)--(2,1)--(2,0);
  \draw[gray!60,dashed] (2,2)--(1,2);
  \draw[Ccol!90!black] (0,2)--(1,2); \node[fill=Ccol,rounded corners=1pt,inner sep=1.3pt] at (0.5,2){\scriptsize $C\,(e_1)$};
  \draw[Dcol!85!black] (1,2)--(2,2); \node[fill=Dcol,rounded corners=1pt,inner sep=1.3pt] at (1.5,2){\scriptsize $D\,(e_2)$};
  \draw[Acol!85!black] (0,1)--(1,1); \node[fill=Acol,rounded corners=1pt,inner sep=1.3pt] at (0.5,1){\scriptsize $A\,(e_3)$};
  \draw[Dcol!85!black] (1,1)--(2,1); \node[fill=Dcol,rounded corners=1pt,inner sep=1.3pt] at (1.5,1){\scriptsize $D\,(e_4)$};
  \draw[Ccol!90!black] (0,2)--(0,1); \node[fill=Ccol,rounded corners=1pt,inner sep=1.3pt] at (0,1.5){\scriptsize $C\,(e_5)$};
  \draw[Bcol!80!black] (1,2)--(1,1); \node[fill=Bcol,rounded corners=1pt,inner sep=1.3pt] at (1,1.5){\scriptsize $B\,(e_6)$};
  \draw[Dcol!85!black] (0,1)--(0,0); \node[fill=Dcol,rounded corners=1pt,inner sep=1.3pt] at (0,0.5){\scriptsize $D\,(e_7)$};
  \draw[Dcol!85!black] (1,1)--(1,0); \node[fill=Dcol,rounded corners=1pt,inner sep=1.3pt] at (1,0.5){\scriptsize $D\,(e_8)$};
  \draw[black,line width=0.8pt] (0.5,1.5) circle[radius=0.7071];
  \draw[black,densely dashed,line width=0.7pt] (0.5,1.5)--(1,2);
  \draw[black,densely dashed,line width=0.7pt] (0.5,1.5)--(1,1);
  \draw[black,densely dashed,line width=0.7pt] (0.5,1.5)--(0,1);
  \node[fill=white,inner sep=0.8pt] at (0.20,1.82){\scriptsize $C$};
  \node[fill=white,inner sep=0.8pt] at (0.82,1.70){\scriptsize $B$};
  \node[fill=white,inner sep=0.8pt] at (0.52,1.25){\scriptsize $A$};
\end{tikzpicture}
\caption{The same partition \eqref{eq:Cpart2}, redrawn to exhibit its
Kitaev--Preskill disk structure. The circle is the boundary of the top-left
plaquette; $A$, $B$, and $C$ are its three arcs ($A=\{e_3\}$ the bottom,
$B=\{e_6\}$ the right, $C=\{e_1,e_5\}$ the top and left), and
$D=\{e_2,e_4,e_7,e_8\}$ is the exterior. Thus $A$, $B$, and $C$ explicitly
form a Kitaev--Preskill disk.}
\label{fig:Appfig4}
\end{figure}

The second step deforms Fig.~\ref{fig:Appfig2} by reassigning edge $e_2$ from
$B$ to $D$,
\begin{equation}
\begin{aligned}
\label{eq:Cpart2}
C=\{e_1,e_5\}, &\quad B=\{e_6\}, \quad  \\ A=\{e_3\}, &\quad D=\{e_2,e_4,e_7,e_8\}.
\end{aligned}
\end{equation}
The corresponding partition is Fig.~\ref{fig:Appfig3}.  This is of
Kitaev--Preskill disk type, as shown in Fig.~\ref{fig:Appfig4}: $A$, $B$, and
$C$ are exactly the three arcs of a single plaquette boundary, with $D$ the
exterior.

\subsection{Invariance of $I_3$ under local deformations}
\label{app:KPdisk-inv}

We now follow the sequence of local deformations from the initial partition
in Fig.~\ref{fig:Appfig1} to the Kitaev--Preskill disk partition in
Fig.~\ref{fig:Appfig4}. At each step, one edge is transferred into the
complementary region $D$, without crossing a non-contractible cycle of the
torus.

For the initial partition \eqref{eq:Cpart0}, direct computation of the
reduced density matrices of $\ket{\psi_{\alpha \beta}}$ gives
\begin{equation}
\begin{aligned}
&S_A=S_B=S_C=S_D=2\log 2,\\
&S_{AB}=S_{BC}=S_{AC}=3\log 2,\quad S_{ABC}=2\log 2,
\end{aligned}
\end{equation}
so that
\begin{equation}
I_3=(6-9+2)\log 2=-\log 2.
\end{equation}
The same calculation can be repeated after each deformation; the results are
collected in Table~\ref{tab:Cdeform} (all entropy entries in units of
$\log 2$).
\begin{table*}[h]
\centering
\caption{Entanglement entropies and tripartite information $I_3$ along the
deformation from the checkerboard partition (Fig.~\ref{fig:Appfig1}) to the
Kitaev--Preskill disk partition (Fig.~\ref{fig:Appfig4}). All entropy entries are in
units of $\log 2$.}
\label{tab:Cdeform}
\begin{tabular}{lccccc}
\hline\hline
$(A;B;C;D)$ & $(|A|,|B|,|C|,|D|)$
& $S_A,S_B,S_C,S_D$ & $S_{AB},S_{BC},S_{AC}$ & $S_{ABC}$ & $I_3/\log 2$ \\
\hline
$e_3e_7;\,e_2e_6;\,e_1e_5;\,e_4e_8$
& $(2,2,2,2)$ & $2,2,2,2$ & $3,3,3$ & $2$ & $-1$ \\
$e_3;\,e_2e_6;\,e_1e_5;\,e_4e_7e_8$
& $(1,2,2,3)$ & $1,2,2,3$ & $3,3,3$ & $3$ & $-1$ \\
$e_3;\,e_6;\,e_1e_5;\,e_2e_4e_7e_8$
& $(1,1,2,4)$ & $1,1,2,3$ & $2,3,3$ & $3$ & $-1$ \\
\hline\hline
\end{tabular}
\end{table*}
Although the individual entropies change at each step, the tripartite
information does not:
\begin{equation}
I_3=-\log 2
\end{equation}
throughout the entire deformation. Since none of these deformations crosses a
non-contractible cycle of the torus, this establishes that $I_3$ is a
topological invariant within this deformation class.

The final partition in Fig.~\ref{fig:Appfig4} is explicitly of
Kitaev--Preskill disk type. For the toric code there are four abelian anyons,
each of quantum dimension $d_i=1$, so the total quantum dimension is
$\mathcal{D}=\sqrt{\sum_i d_i^2}=\sqrt4=2$, and the Kitaev--Preskill
topological entanglement entropy is $\gamma=\log\mathcal{D}$. Since the
partition \eqref{eq:Cpart0} lies in the same topological class as this disk
partition---it can be deformed into it without any region wrapping a
non-contractible cycle---the value $I_3=-\log 2$ obtained above legitimately
measures the topological entanglement entropy:
\begin{equation}
I_3=-\gamma=-\log\mathcal{D}=-\log 2.
\end{equation}

\end{document}